\documentclass[authoryear]{elsarticle}
\usepackage{epsfig,graphicx,aas_macros}
\usepackage{graphics}
\usepackage{amsmath}
\usepackage[font=small,labelfont=bf]{caption}
\usepackage[authoryear]{natbib}
\begin{document}

\begin{frontmatter}

\title{Temperature Measurement during Thermonuclear X-ray Bursts with \emph{BeppoSAX}}
\author[mymainaddress]{Aru Beri \corref{mycorrespondingauthor}}
\ead{aruberi@iitrpr.ac.in}
\cortext[mycorrespondingauthor]{Corresponding author}
\address[mymainaddress]{Department of Physics, Indian Institute of Technology Ropar, Nangal Road, Rupnagar, Punjab, 140001 India}
\author[Raman]{Biswajit~Paul}
\address[Raman]{Raman Research Institute, C. V. Raman Avenue, Sadashivanagar, Bangalore-560012, India }
\author[inaf]{Mauro~Orlandini}
\address[inaf]{INAF/IASF-Bologna, via Gobetti 101, I-40129 Bologna, Italy.}
\author[france]{Chandreyee~Maitra}
\address[france]{Laboratoire AIM, CEA-IRFU/CNRS/Universit ́e Paris Diderot, Service d’Astrophysique, CEA Saclay, F-91191 Gif sur Yvette, France}

\begin{abstract}

 We have carried out a study of temperature evolution during thermonuclear bursts in LMXBs
using broad band data from two instruments onboard \emph{BeppoSAX}, the \textsc{MECS}
and the \textsc{PDS}. However, instead of applying the standard technique of time resolved spectroscopy,
we have determined the temperature in small time intervals using the ratio of 
count rates in the two instruments assuming a blackbody nature of burst emission
and different interstellar absorption for different sources. Data from a total
of twelve observations of six sources were analysed during which 22 bursts
were detected. We have obtained temperatures as high
as $\sim$3.0~keV, even when there is no evidence of photospheric radius expansion.
These high temperatures were observed in the sources within different broadband spectral states (soft and hard).
\end{abstract}


\end{frontmatter}
\section{INTRODUCTION}

Thermonuclear X-ray bursts due to unstable nuclear burning of hydrogen and/or helium \citep{Joss77, Lamb78,
Lewin93,Lewin95, Strohmayer06, Bhattacharyya10}
have been observed in nearly 80 neutron star low mass X-ray binaries 
\citep{Liu07, Galloway08}.
These bursts offer a useful tool for the measurement of neutron star parameters \citep{Bhattacharyya10}.
Time resolved spectroscopy during bursts have been performed for many sources for the determination
of neutron star radius by assuming that the entire surface emits in X-rays 
\citep[e.g.,][]{vanParadijs78, Galloway08, Guver12a}.
The background subtracted continuum spectra during these bursts are often fit using Planck (blackbody)
function. The persistent emission prior to the burst is subtracted as background \citep{Galloway08, Bhattacharyya10}.
In time resolved spectroscopy of the photospheric radius expansion bursts, when the
photosphere falls back to the neutron star surface, the temperature has the
highest value and the blackbody normalisation has the lowest value, which is also
called the touchdown \citep[e.g.,][]{Damen90, Kuulkers03}.
Time resolved spectroscopy during the cooling phase after the touchdown is the
most widely used method for neutron star radius measurement 
\citep[e.g.,][]{Lewin93, Ozel06, Galloway08, Ozel09, Guver10a, Guver10b, Guver12a, Guver12b}.
The scattering of photons by the electrons and frequency dependence of the
opacity in the neutron star atmosphere harden the spectrum, and shift it
to higher energies \citep{London84, London86, Syunyaev86, Ebisuzaki88,Titarchuk94, Madej04, Majczyna05, Bhattacharyya10}.
Therefore, it is believed that the effective temperature is substantially smaller than the
temperature obtained from the blackbody fit \citep[e.g.,][]{Ebisuzaki84, Galloway08}.
The observed color temperature and flux is associated
with the blackbody radius through 
$R_{\infty}=(F_{\infty}/{\sigma}T_{\infty}^4)^{1/2}d$ 
\citep{Lewin93}. Here, $F_{\infty}$ is the observed flux, $T_{\infty}$
is the blackbody temperature measured at infinity and $d$ refers to the source distance.
The neutron star radius is estimated from the 
blackbody radius ($R_{BB}$) via the following equation. 
\begin{equation}
 R_{BB}=R_{\infty}{f_{c}}^{2}/(1+z)
\end{equation}
where, $z$ is the gravitational redshift and $f_c$ is the color correction factor
which is defined as the ratio of color temperature ($T_c$) and the effective
temperature ($T_{eff}$) of the star \citep{London86, Madej04, Majczyna05, Suleimanov11a,
 Suleimanov12}. Color correction factor varies as a function of effective temperature and also depends on
effective gravitational acceleration, which determines the density profiles
of the atmospheric layers \citep{Guver12a}. It has been found that X-ray burst cooling properties 
are dependent on the accretion rate and the spectral states \citep{Suleimanov11b, Kajava14}.
Though a color correction factor close to 1.4 has often been used \citep{Madej04, Majczyna05},
the most recent calculations by \citet{Suleimanov12}  
suggest that $f_c$ is in range of 1.8-1.9 when the luminosity is 
close to Eddington luminosity~($L_{Edd}$) and its value decreases to a range of 1.4-1.5 with
the subsequent fall to $\sim$ 0.5 $L_{Edd}$. 
Assuming a constant value of color correction factor may lead to systematic change 
in inferred apparent surface area \citep{Guver12a}. 
 Hence, this is one of the sources for systematic
uncertainity while measuring the radius of a neutron star from X-ray bursts. \\

Even if a blackbody model 
provides a good fit for the time resolved burst spectrum that is often measured with the proportional counters, 
\citet{Nakamura89} have reported deviations from a
blackbody. The authors observed a high energy tail during bursts 
and have interpreted it as a result of comptonization of the burst emission
by hot plasma surrounding the neutron star.
%
For a peak temperature close to 2.5~keV, the emission peaks at 2.8$\times$kT, i.e 6-7~keV.
Therefore, in addition to the \emph{RXTE}-\textsc{PCA} we expect to detect the bursts even with high energy instruments like 
\emph{BeppoSAX}-\textsc{PDS}, \emph{Suzaku}-\textsc{PIN}, \emph{NuSTAR}. However, simultaneous data at
energies below 10~keV with sufficient time resolution is also required to measure the temperature evolution and
this is possible with the \textsc{MECS} of \emph{BeppoSAX} and also with \emph{NuSTAR}. 
\citet{Barriere14}, using \emph{NuSTAR} data have reported a Type-I burst in GRS~1741.9--2853 that was 
found to be 800~seconds long with mild Photospheric radius expansion (PRE).
The peak temperature of this burst was found to be 2.65$\pm$0.06~keV.  \\


A large fraction of all the burst temperature studies have been done with \emph{RXTE-PCA} in the energy range of 3-25~keV.
We have estimated the temperature evolution in short intervals during bursts
that were observed with \emph{BeppoSAX}. 
We performed studies of bursts using the two instruments \textsc{MECS}
and \textsc{PDS} on-board \emph{BeppoSAX}. Since \emph{BeppoSAX} data
have lower count rates than \emph{RXTE-PCA} we have used a new technique for measuring the
temperature evolution.
If a blackbody spectrum with a fixed absorption column density is fitted to the time
resolved spectra, the temperature obtained is a function of the ratio of the count rates
in two energy bands. Therefore, instead of a spectral fit of data with low statistical
quality, we have used the hardness ratio (HR) to determine the temperature evolution. 
The paper is organised as : 
the second section describes the observations and data reduction 
procedure, the third and fourth sections describe the calibration and timing studies performed.
The last section is dedicated to the implications of the results achieved.

\section{OBSERVATIONS AND DATA REDUCTION}
\emph{BeppoSAX} had four co-aligned 
narrow field instruments (NFI) \citep{Boella97} and a Wide Field Camera (\textsc{WFC})~\citep{Jager97}.
The four NFIs are:
i)~The Medium-Energy Concentrator Spectrometer (\textsc{MECS}) that consists of three grazing incidence telescopes each with 
an imaging gas proportional counter that work in 1.3-10~keV band \citep{Boella97},
ii)~The Low-Energy Concentrator Spectrometer (\textsc{LECS}), consisting of similar kind of imaging gas scintillation 
proportional counters but with an ultra-thin (1.25~$\mu$m) entrance window and working in the energy range of
0.1-10~keV \citep{Parmar97}, iii)~The High Pressure Gas Scintillation proportional Counter 
(\textsc{HPGSPC},~4-120~keV; \citet{Manzo97})
and iv)~The Phoswich Detection System (\textsc{PDS},~15-300~keV; \citet{Frontera97}).
The \textsc{HPGSPC} and \textsc{PDS} are non-imaging instruments.
The \textsc{PDS} detector is composed of 4 actively shielded NaI(Tl)/CsI(Na) phoswich scintillators 
with a total geometric area of 795 $cm^2$. \\

Bursts were clearly detected over the entire energy range of 1.8-10~keV of \textsc{MECS}
while in the case of \textsc{PDS}, most of the bursts were noticable only upto 30~keV, i.e. in the energy
range of 15-30~keV. Therefore, we selected these two energy bands for estimating
the hardness ratio.
%

We considered only those sources and observations for which bursts were observed 
simultaneously in \textsc{MECS} and \textsc{PDS}. We have found a total of 22 bursts in 6 sources.
The log of observations is given in Table-1. Since \textsc{MECS~2} and \textsc{MECS~3} data were available for all 
sources considered, we merged data from both these MECS. 
\emph{HEASOFT-6.12} and \emph{SAXDAS} (version 2.3.3) were used for reduction
and extraction purposes. \\

Subsequently, the merged \textsc{MECS} event data files were used for extraction of light curves
with a binsize of 0.5~seconds using the ftools \footnote{http://heasarc.gsfc.nasa.gov/ftools/} task \emph{xselect}.
The source radius of $4^{\prime}$ corresponding to $\simeq$95 $\%$ of the 
instrumental Point Spread Funcions was selected and appropriate good time intervals~(GTI) 
were applied.
The light curves were restricted to the energy band 1.8-10~keV using appropriate energy filters.
In case of \textsc{PDS}, the \textsc{SAXDAS} programs \emph{saxpipe} and \emph{pdproducts} were used
for creating the light curves with a bin time of 0.5~seconds. 
Figure-\ref{lcurves} shows the time series for all the sources including about 100 secs of data before
and after the bursts. It is evident from the light curves shown in Figure-\ref{lcurves}
that the persistent emission is stable before and after the burst.
An interesting feature seen in one of the sources is the double peaked behaviour of the burst profile from \emph{PDS} data
of the source SAX~J1747--2853 (Observation ID-210320013). However, this feature was not seen
in the light curve created using \textsc{MECS} data. The burst profile of the same source created
using the \textsc{WFC} data also showed a prominent double peaked behaviour in 
the high energy band (8-28)~keV \citep{Natalucci00}.
Similar kind of bi-horned profile from the X-ray burst in the higher energy band (\textsc{PDS}) has also been observed in X~1724-308. \\

We subtracted the average pre and post burst count rates to obtain only the burst profile
in the two energy bands, namely 1.3-10~keV and 15.0-30.0~keV.
These burst profiles were then used to calculate the Hardness Ratio~(HR) which is
defined as ratio of count rates between two instruments \emph{PDS} 
and \textsc{MECS} near the peak of these bursts (see, Figure-\ref{HR}). \\ 

A burst in 4U~1702--429 showed the maximum value of the hardness ratio of the order of $\sim$0.9$\pm$0.1.
The three bursts in MXB~1728--34 reached a value upto $\sim$ 0.7$\pm$0.1 in the hardness.
The burst obtained from the observation (ID-210320013)
of SAX~J1747--2853 in which \textsc{PDS} profile showed a double peak behaviour,
the highest value seen was $\sim$0.95$\pm$0.18.
However, the maximum value was close to 0.5$\pm$0.1 in the burst obtained from the other observation (ID-21032001)
of SAX~J1747--2853. 
For the bursts from the sources namely, SAX~J1748.9--202, X~1724--308
and GS~1826--238, the highest values obtained for hardness ratio was close to 0.3$\pm$0.1.

\begin{table}
\centering
\caption{Log of observations used in this work}
\begin{tabular}{@{}lccccc@{}}
\hline
\hline

Source Name           & Obs-ID      & Observation Date &       & Number of Bursts                       & $N_H$         \\
                     &        &                  &                     &                               & ($10^{22}$) $cm^2$   \\
\hline

4U~1702--429        & 21224001         & 2000-08-24  &                & 3          & $1.8^b$     \\
4U~1702--429                 & 21224002  & 2000-09-23 &                 & 2          & $1.8^b$     \\
X~1724--308                    & 20105002 & 1996-08-17 &                & 1          &  $1.11^e$   \\
4U~1728--34                  & 20674001  & 1998-08-23 &               & 3          & $2.5^a$  \\
SAX~J1747--2853              & 21032001   & 2000-03-16 &              & 1          & $8.8^c$     \\
SAX~J1747--2853               & 210320013 & 2000-04-12 &               & 1          &  $8.8^c$     \\
SAX~J1748.9--202             & 20549003   & 1998-08-26 &              & 1          &  $0.82^d$     \\
SAX~J1748.9--202            & 21416001   & 2001-10-02 &              & 3          &  $0.82^d$     \\

GS~1826--238                  & 20263003    & 1997-10-25 &             & 1          &  $0.11^f$      \\
GS~1826--238                & 21024001       & 1999-10-20 &          & 1          &  $0.11^f$     \\
GS~1826--238                 & 21024002       & 2000-04-18 &          & 3          &  $0.11^f$   \\
GS~1826--238               & 20269001         & 1997-04-06 &        & 2          &  $0.11^f$    \\

\hline
\end{tabular}

\bigskip

{\bf{References}}: $N_H$ values were taken from: \\
             a) \citet{Salvo00} b) \citet{Church14}, c) \citet{Natalucci04} 
             d) \citet{Zand99b} e) \citet{Guainazzi98} f) \citet{Zand99a}   \\

\end{table}

\section{Temperature Measurements}
\begin{figure*}
\begin{minipage}{0.35\textwidth}
\includegraphics[height=3.5in,width=3.0in,angle=0]{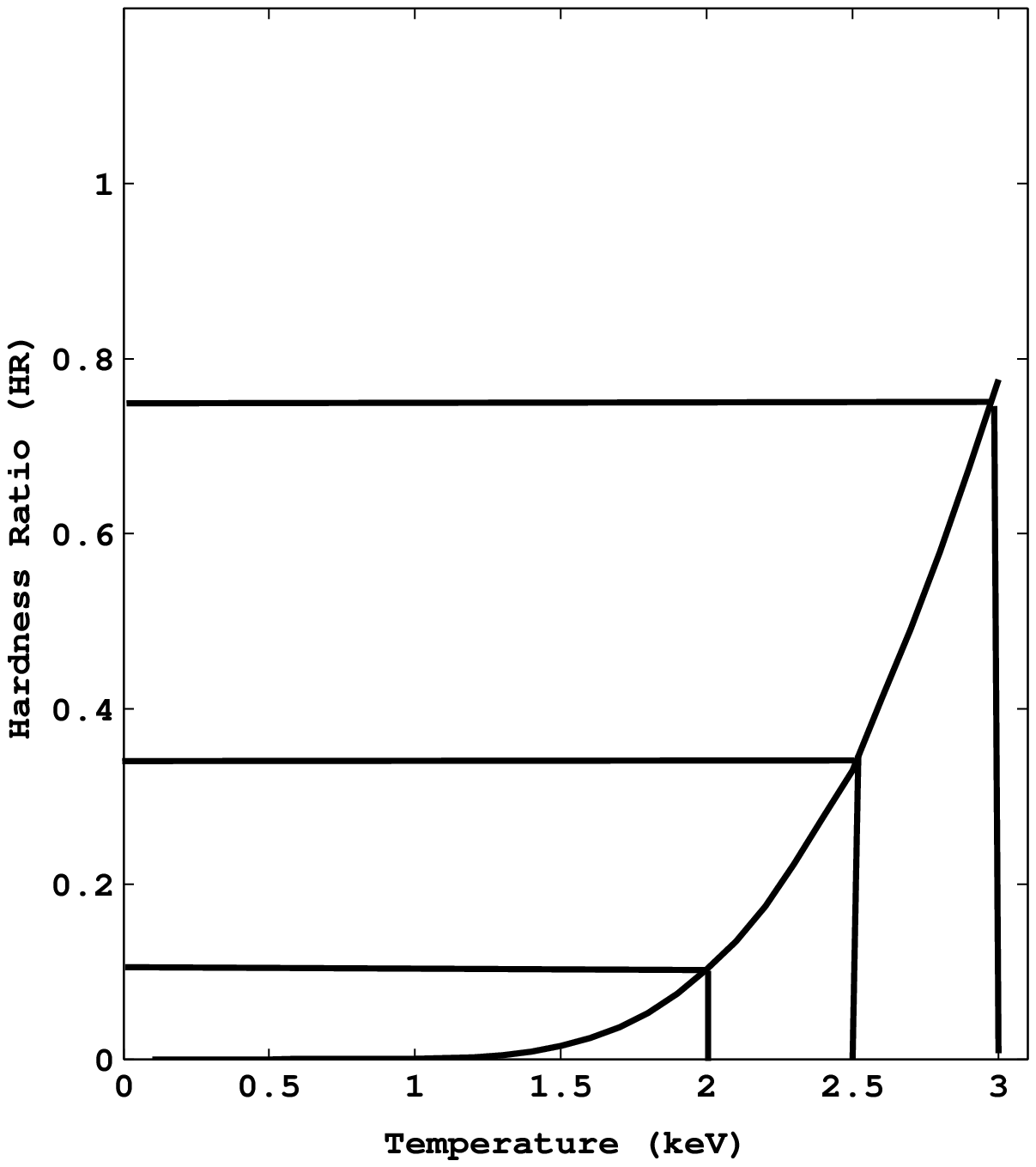}
\caption{Temperature versus Hardness Ratio curve obtained using $N_H$=0.1$\times{10^{22}}$.
 In this plot, three horizontal lines indicates the ratio values for three different temperatures 2~keV, 2.5~keV and 3~keV.}
 \label{T_HR}
 \end{minipage}
 \hspace{0.3\linewidth}
 \begin{minipage}{0.4\textwidth}
 
\includegraphics[height=4.in,width=3.in,angle=0]{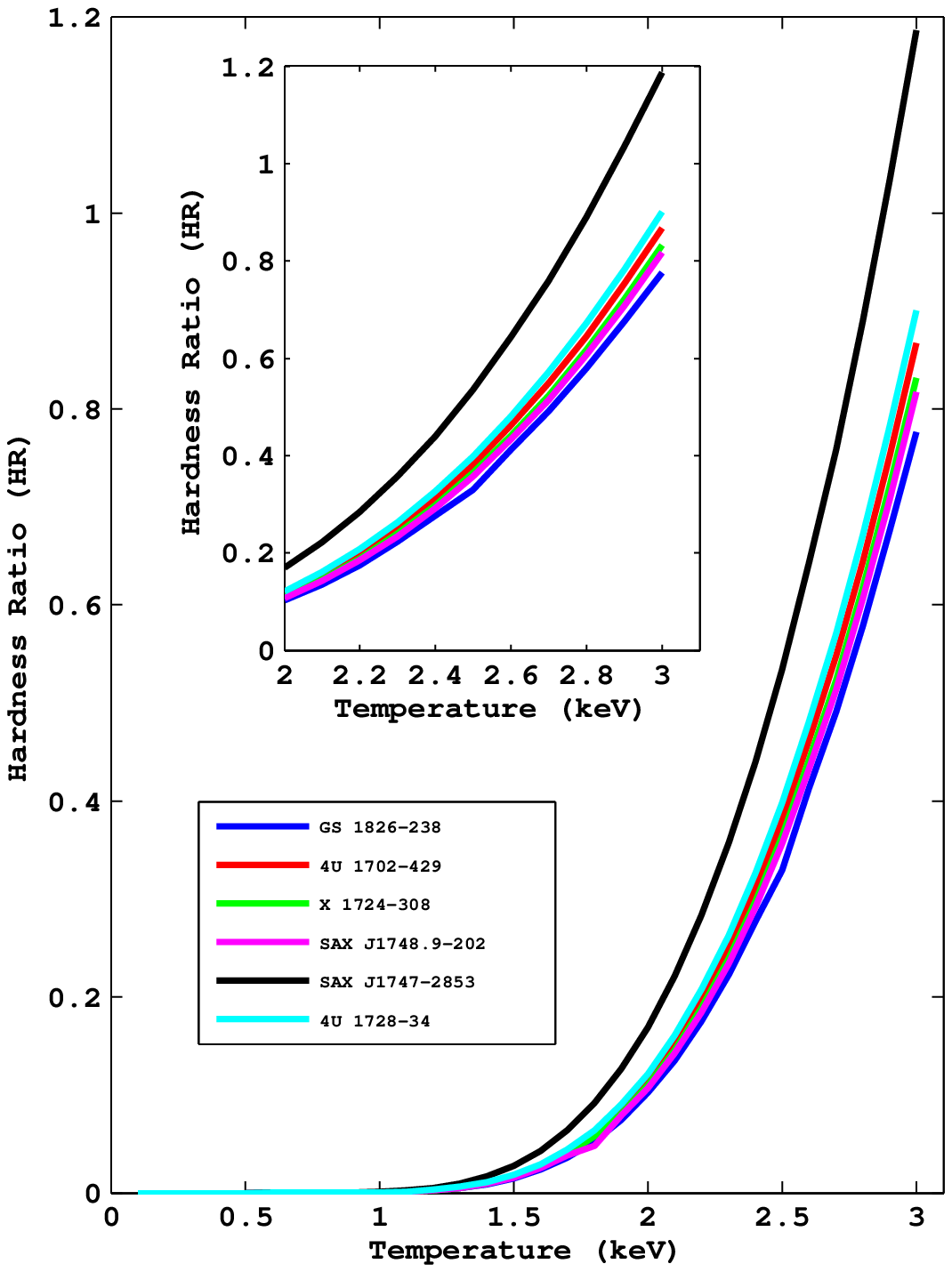}
 \caption{Temperature versus Hardness Ratio calibration curves obtained using known values
 of interstellar absorption for different sources. In this plot, each color correspond to
 a curve for a specific source as marked.}
 \label{T_HR_NH}
 \end{minipage}
 \end{figure*}

Subsequently, after finding the values of hardness ratios during different bursts
in various sources, we aim at finding the corresponding temperatures.
Using the assumption that during thermonuclear bursts, neutron star emits like a blackbody,
the hardness ratio, in principle  can be directly converted to a corresponding blackbody temperature.
We simulated the absorbed blackbody spectrum using \emph{XSPEC}.
The response files of \textsc{MECS~2}, \textsc{MECS~3} and \textsc{PDS} released by 
\emph{BeppoSAX} Science Data Center (SDC) in 1997 along with corresponding ancillary files were used
for this purpose. The simulated spectra at different temperatures 
were used to calculate the expected hardness ratio between two instruments
at different blackbody temperatures for $N_H$ values appropriate for different sources. \\
For the simulation we have used the
\texttt{bbodyrad model }multiplied with interstellar 
absorption component \texttt{phabs} which is available as a standard model
in \emph{XSPEC}\footnote{https://heasarc.gsfc.nasa.gov/xanadu/xspec/manual/XSmodelPhabs.html} . 
The spectrum was simulated for temperature
values between 0.1-3~keV. Initially, we started with hydrogen column density ($N_H$)=0.1$\times{10^{22}}$
for obtaining the Temperature~(T)~versus Hardness~ratio~(HR) curve (see Figure-\ref{T_HR}).
Miller, Cackett, and Reis (2009) using high resolution grating spectra 
showed that individual photoelectric absorption edges
observed in X-ray spectra of a number of X-ray binaries are independent of their spectral states.
Thus, one could fix the values of absorption in the interstellar medium to the previously 
known values. For this reason we have fixed the values of $N_{H}$ to those of the
the non-burst spectra from the same set of observations.
These values are given in Table-1.
Using these respective values for each source
we obtained the values of temperature corresponding to
their hardness ratio (see Figure-\ref{T_HR_NH}).
Next, using these simulated curves we measured the temperature evolution of all the bursts~(Figure-\ref{Temperature}).
Spline interpolation was used for the 
estimation of the temperatures. Errors on the temperature were estimated by propogating errors
on the count-rates. \\
From Figure-\ref{Temperature}, it is interesting to notice that the bursts in 4U~1702--429
with the maximum HR ratio values
close to $\sim$0.9$\pm$0.1 showed a temperature as high as 3.0$\pm$0.1~keV.
Here, we would like to  mention that \citet{Guver12a} discussed the bursts from 4U~1702--429 
observed with \emph{RXTE} (not the same bursts reported here from \emph{BeppoSAX})
and found that a blackbody does not provide a good
fit for these bursts which have low peak flux. 
The maximum temperature seen in the bursts from 4U~1728--34 was $\sim$~2.7$\pm$0.1~keV.
The burst from the source SAX~J1747--2853, that showed
a double peaked burst profile in PDS exhibited a temperature upto 2.7$\pm$0.1~keV
while the remaining bursts had highest temperatures close to 2.2$\pm$0.1~keV.
The occurrence of high temperatures in some of the bursts was in agreement 
with those observed with \emph{RXTE}-PCA \citep[see, e.g.,][]{Boutloukos10}. \\
We further investigated the spectral states of the sources that showed temperatures
in the range of 2.5-3.0~keV. 
During the \emph{BeppoSAX} observations 4U~1728--34 \citep{Salvo00} and SAX~J1748.9--202 \citep{Zand99b}
were in soft spectral state while 4U~1702--429 \citep{Church14}, SAX~J1747--2853 \citep{Natalucci04},
GS~1826--238 \citep{Zand99a,Sordo99,Cocchi01,Cocchi11}, X~1724--308 \citep{Guainazzi98}
were in hard spectral state.
An interesting feature to note from all the studies carried out by different authors
using the narrow field instruments (NFI) on-board
\emph{BeppoSAX} was that the persistent emission was modelled as comptonized spectrum. 

\section{Summary and Discussions}
We have presented an analysis of 22 X-ray bursts from 6 LMXB systems using the 
data from the two instruments \textsc{MECS} and \textsc{PDS} onboard \emph{BeppoSAX}. \\

The maximum allowed value of effective temperature 
for Eddington limited luminosity is expected to be close to 1.7~keV for a neutron star
surrounded by an atmosphere of fully ionised hydrogen \citep{Lewin93}.
If the effective temperature exceeds 2.0~keV for a neutron star
surrounded by an atmosphere of fully ionised helium, the radiative flux is greater than
Eddington flux \citep{Boutloukos10}. Hence an effective temperature greater than 2~keV is not expected
in the thermonuclear bursts. 
However, we have found that 
temperatures attain values as high as $\sim$3.0~keV in some of the bursts. \\ 

It is believed that bursts occuring during different states (hard or soft)
exhibit a different behaviour.
During a soft state, bursts do not follow 
theoretically predicted NS atmospheric models \citep{Poutanen14, Kajava14}. 
To our surprise, it was found that the two sources 4U~1702-429 and 4U~1728--34 that 
showed temperatures greater than 2.7~keV were in different spectral states,
4U~1702-429, in a hard state while 4U~1728--34 was in a soft state. 
Two spectral states are believed to occur because of a change in the accretion geometry \citep{Kajava14}.
It is even more difficult to understand a large effective temperature of some of the
bursts in two different spectral states, when the comptonisation effects would be
different. We note that usefulness for radius measurement of the X-ray bursts in soft state
has already been questioned \citep{Poutanen14, Kajava14}. \\

A considerable contribution from comptonisation
of the persistent emission or a deviation from a true blackbody spectrum
may lead to the high temperatures measured in the case of the burst from SAX~J1747--2853.
This is also quite evident from the presence of a double peaked burst profile 
only in \textsc{PDS} data. Alternatively,
there also seems to be a type of bursts where
the color temperature exceeds 2.5~keV in the same way as expected from PRE bursts but the corresponding
normalisation does not vary much \citep[eg.][]{Kaptein00, Van01}.
This kind of peculiar behaviour is believed to be due to sudden change in the color
correction factor. \citet{Boutloukos10} have also discussed the temperatures greater than 2~keV during bursts with no evidence of
radius expansion in them, implying
super-Eddington fluxes.
The same authors suggest that these high temperatures could be due to 
comptonization of the surface emission that alter its spectrum
as well as the relation between energy flux and apparent surface
area. The thermonuclear bursts detected with \emph{BeppoSAX} provide
evidence of a significantly higher temperature of the bursts than expected or a  significant 
deviation from blackbody spectrum, perhaps by the process of comptonisation
, which is evident in the persistent emission during the same observations. \\

%


\section*{Acknowledgments}
A.B would like to thank, Bryan.K.Ibry (NASA) and Jacob~Rajan (RRI)
for helping out during the installation of \emph{SAXDAS} software.
The research has made use of data obtained from High Energy Astrophysics
Science Archive Research Center (\emph{HEASARC}) and ASI Science Data Center.
A.B also acknowledges IIT Ropar for financial assistance and Raman
Research Institute (RRI) for providing local hospitality.



\bibliographystyle{elsarticle-harv}

 \bibliography{references}

\begin{thebibliography}{50}
\expandafter\ifx\csname natexlab\endcsname\relax\def\natexlab#1{#1}\fi
\expandafter\ifx\csname url\endcsname\relax
  \def\url#1{\texttt{#1}}\fi
\expandafter\ifx\csname urlprefix\endcsname\relax\def\urlprefix{URL }\fi

\bibitem[{{Barri{\`e}re} et~al.(2014){Barri{\`e}re}, {Tomsick}, {Baganoff},
  {Boggs}, {Christensen}, {Craig}, {Dexter}, {Grefenstette}, {Hailey},
  {Harrison}, {Madsen}, {Mori}, {Stern}, {Zhang}, {Zhang}, and
  {Zoglauer}}]{Barriere14}
{Barri{\`e}re}, N.~M., {Tomsick}, J.~A., {Baganoff}, F.~K., {Boggs}, S.~E.,
  {Christensen}, F.~E., {Craig}, W.~W., {Dexter}, J., {Grefenstette}, B.,
  {Hailey}, C.~J., {Harrison}, F.~A., {Madsen}, K.~K., {Mori}, K., {Stern}, D.,
  {Zhang}, W.~W., {Zhang}, S., {Zoglauer}, A., May 2014. {NuSTAR Detection of
  High-energy X-Ray Emission and Rapid Variability from Sagittarius
  A$^{sstarf}$ Flares}. \apj 786, 46.

\bibitem[{{Bhattacharyya}(2010)}]{Bhattacharyya10}
{Bhattacharyya}, S., Apr. 2010. {Measurement of neutron star parameters: A
  review of methods for low-mass X-ray binaries}. Advances in Space Research
  45, 949--978.

\bibitem[{{Boella} et~al.(1997){Boella}, {Chiappetti}, {Conti}, {Cusumano},
  {del Sordo}, {La Rosa}, {Maccarone}, {Mineo}, {Molendi}, {Re}, {Sacco}, and
  {Tripiciano}}]{Boella97}
{Boella}, G., {Chiappetti}, L., {Conti}, G., {Cusumano}, G., {del Sordo}, S.,
  {La Rosa}, G., {Maccarone}, M.~C., {Mineo}, T., {Molendi}, S., {Re}, S.,
  {Sacco}, B., {Tripiciano}, M., Apr. 1997. {The medium-energy concentrator
  spectrometer on board the BeppoSAX X-ray astronomy satellite}. \aaps 122,
  327--340.

\bibitem[{{Boutloukos} et~al.(2010){Boutloukos}, {Miller}, and
  {Lamb}}]{Boutloukos10}
{Boutloukos}, S., {Miller}, M.~C., {Lamb}, F.~K., Sep. 2010. {Super-Eddington
  Fluxes During Thermonuclear X-ray Bursts}. \apjl 720, L15--L19.

\bibitem[{{Church} et~al.(2014){Church}, {Gibiec}, and
  {Ba{\l}uci{\'n}ska-Church}}]{Church14}
{Church}, M.~J., {Gibiec}, A., {Ba{\l}uci{\'n}ska-Church}, M., 2014. {The
  nature of the island and banana states in atoll sources and a unified model
  for low-mass X-ray binaries}. \mnras 438, 2784.

\bibitem[{{Cocchi} et~al.(2001){Cocchi}, {Bazzano}, {Natalucci}, {Ubertini},
  {Heise}, {Kuulkers}, and {in't Zand}}]{Cocchi01}
{Cocchi}, M., {Bazzano}, A., {Natalucci}, L., {Ubertini}, P., {Heise}, J.,
  {Kuulkers}, E., {in't Zand}, J.~J.~M., Jan. 2001. {Beppo-SAX observation of
  the burster GS 1826-238}. Advances in Space Research 28, 375--379.

\bibitem[{{Cocchi} et~al.(2011){Cocchi}, {Farinelli}, and {Paizis}}]{Cocchi11}
{Cocchi}, M., {Farinelli}, R., {Paizis}, A., May 2011. {BeppoSAX view of the
  NS-LMXB GS 1826-238}. \aap 529, A155.

\bibitem[{{Damen} et~al.(1989){Damen}, {Jansen}, {Penninx}, {Oosterbroek}, {van
  Paradijs}, and {Lewin}}]{Damen90}
{Damen}, E., {Jansen}, F., {Penninx}, W., {Oosterbroek}, T., {van Paradijs},
  J., {Lewin}, H.~G., Mar. 1989. {Non-Planckian behaviour of burst spectra -
  Dependence of the blackbody radius on the duration of bursts}. \mnras 237,
  523--531.

\bibitem[{{del Sordo} et~al.(1999){del Sordo}, {Frontera}, {Pian}, {Piraino},
  {Oosterbroek}, {Harmon}, {Palazzi}, {Tavani}, {Zhang}, and
  {Parmar}}]{Sordo99}
{del Sordo}, S., {Frontera}, F., {Pian}, E., {Piraino}, S., {Oosterbroek}, T.,
  {Harmon}, B.~A., {Palazzi}, E., {Tavani}, M., {Zhang}, S.~N., {Parmar},
  A.~N., 1999. {BeppoSAX Observations of the Galactic Source GS 1826-238 in a
  Hard X-Ray High State}. Astrophysical Letters and Communications 38, 125.

\bibitem[{{Di Salvo} et~al.(2000){Di Salvo}, {Iaria}, {Burderi}, and
  {Robba}}]{Salvo00}
{Di Salvo}, T., {Iaria}, R., {Burderi}, L., {Robba}, N.~R., Oct. 2000. {The
  Broadband Spectrum of MXB 1728-34 Observed by BeppoSAX}. \apj 542,
  1034--1040.

\bibitem[{{Ebisuzaki} and {Nakamura}(1988)}]{Ebisuzaki88}
{Ebisuzaki}, T., {Nakamura}, N., May 1988. {The difference in hydrogen
  abundance between two classes of type I X-ray bursts}. \apj 328, 251--255.

\bibitem[{{Ebisuzaki} et~al.(1984){Ebisuzaki}, {Sugimoto}, and
  {Hanawa}}]{Ebisuzaki84}
{Ebisuzaki}, T., {Sugimoto}, D., {Hanawa}, T., 1984. {Are X-ray bursts really
  of super-Eddington luminosities?} \pasj 36, 551--566.

\bibitem[{{Frontera} et~al.(1997){Frontera}, {Cinti}, {Dal Fiume}, {Landini},
  {Nicastro}, {Orlandini}, {Zavattini}, {Costa}, {Schreiner}, {Rosza}, {Raby},
  {White}, {Chiaverini}, {Monzani}, {Poulsen}, and {Suetta}}]{Frontera97}
{Frontera}, F., {Cinti}, M.~N., {Dal Fiume}, D., {Landini}, G., {Nicastro}, L.,
  {Orlandini}, M., {Zavattini}, G., {Costa}, E., {Schreiner}, R.~S., {Rosza},
  C.~M., {Raby}, P.~S., {White}, J., {Chiaverini}, V., {Monzani}, F.,
  {Poulsen}, J.~M., {Suetta}, E., Oct. 1997. {On-ground performance tests of
  the SAX/PDS detector.} Nuovo Cimento C Geophysics Space Physics C 20,
  797--809.

\bibitem[{{Galloway} et~al.(2008){Galloway}, {Muno}, {Hartman}, {Psaltis}, and
  {Chakrabarty}}]{Galloway08}
{Galloway}, D.~K., {Muno}, M.~P., {Hartman}, J.~M., {Psaltis}, D.,
  {Chakrabarty}, D., Dec. 2008. {Thermonuclear (Type I) X-Ray Bursts Observed
  by the Rossi X-Ray Timing Explorer}. \apjs 179, 360--422.

\bibitem[{{Guainazzi} et~al.(1998){Guainazzi}, {Parmar}, {Segreto}, {Stella},
  {dal Fiume}, and {Oosterbroek}}]{Guainazzi98}
{Guainazzi}, M., {Parmar}, A.~N., {Segreto}, A., {Stella}, L., {dal Fiume}, D.,
  {Oosterbroek}, T., Nov. 1998. {The comptonized X-ray source X 1724-308 in the
  globular cluster Terzan 2}. \aap 339, 802--810.

\bibitem[{{G{\"u}ver} et~al.(2010{\natexlab{a}}){G{\"u}ver}, {{\"O}zel},
  {Cabrera-Lavers}, and {Wroblewski}}]{Guver10a}
{G{\"u}ver}, T., {{\"O}zel}, F., {Cabrera-Lavers}, A., {Wroblewski}, P., Apr.
  2010{\natexlab{a}}. {The Distance, Mass, and Radius of the Neutron Star in 4U
  1608-52}. \apj 712, 964--973.

\bibitem[{{G{\"u}ver} et~al.(2012{\natexlab{a}}){G{\"u}ver}, {{\"O}zel}, and
  {Psaltis}}]{Guver12b}
{G{\"u}ver}, T., {{\"O}zel}, F., {Psaltis}, D., Mar. 2012{\natexlab{a}}.
  {Systematic Uncertainties in the Spectroscopic Measurements of Neutron-star
  Masses and Radii from Thermonuclear X-Ray Bursts. II. Eddington Limit}. \apj
  747, 77.

\bibitem[{{G{\"u}ver} et~al.(2012{\natexlab{b}}){G{\"u}ver}, {Psaltis}, and
  {{\"O}zel}}]{Guver12a}
{G{\"u}ver}, T., {Psaltis}, D., {{\"O}zel}, F., Mar. 2012{\natexlab{b}}.
  {Systematic Uncertainties in the Spectroscopic Measurements of Neutron-star
  Masses and Radii from Thermonuclear X-Ray Bursts. I. Apparent Radii}. \apj
  747, 76.

\bibitem[{{G{\"u}ver} et~al.(2010{\natexlab{b}}){G{\"u}ver}, {Wroblewski},
  {Camarota}, and {{\"O}zel}}]{Guver10b}
{G{\"u}ver}, T., {Wroblewski}, P., {Camarota}, L., {{\"O}zel}, F., Aug.
  2010{\natexlab{b}}. {The Mass and Radius of the Neutron Star in 4U 1820-30}.
  \apj 719, 1807--1812.

\bibitem[{{in 't Zand} et~al.(1999{\natexlab{a}}){in 't Zand}, {Heise},
  {Kuulkers}, {Bazzano}, {Cocchi}, and {Ubertini}}]{Zand99a}
{in 't Zand}, J.~J.~M., {Heise}, J., {Kuulkers}, E., {Bazzano}, A., {Cocchi},
  M., {Ubertini}, P., Jul. 1999{\natexlab{a}}. {Broad-band X-ray measurements
  of GS 1826-238}. \aap 347, 891--896.

\bibitem[{{in 't Zand} et~al.(1999{\natexlab{b}}){in 't Zand}, {Verbunt},
  {Strohmayer}, {Bazzano}, {Cocchi}, {Heise}, {van Kerkwijk}, {Muller},
  {Natalucci}, {Smith}, and {Ubertini}}]{Zand99b}
{in 't Zand}, J.~J.~M., {Verbunt}, F., {Strohmayer}, T.~E., {Bazzano}, A.,
  {Cocchi}, M., {Heise}, J., {van Kerkwijk}, M.~H., {Muller}, J.~M.,
  {Natalucci}, L., {Smith}, M.~J.~S., {Ubertini}, P., May 1999{\natexlab{b}}.
  {A new X-ray outburst in the globular cluster NGC 6440: SAX J1748.9-2021}.
  \aap 345, 100--108.

\bibitem[{{Jager} et~al.(1997){Jager}, {Mels}, {Brinkman}, {Galama},
  {Goulooze}, {Heise}, {Lowes}, {Muller}, {Naber}, {Rook}, {Schuurhof},
  {Schuurmans}, and {Wiersma}}]{Jager97}
{Jager}, R., {Mels}, W.~A., {Brinkman}, A.~C., {Galama}, M.~Y., {Goulooze}, H.,
  {Heise}, J., {Lowes}, P., {Muller}, J.~M., {Naber}, A., {Rook}, A.,
  {Schuurhof}, R., {Schuurmans}, J.~J., {Wiersma}, G., Nov. 1997. {The Wide
  Field Cameras onboard the BeppoSAX X-ray Astronomy Satellite}. \aaps 125,
  557--572.

\bibitem[{{Joss}(1977)}]{Joss77}
{Joss}, P.~C., Nov. 1977. {X-ray bursts and neutron-star thermonuclear
  flashes}. \nat 270, 310--314.

\bibitem[{{Kajava} et~al.(2014){Kajava}, {N{\"a}ttil{\"a}}, {Latvala},
  {Pursiainen}, {Poutanen}, {Suleimanov}, {Revnivtsev}, {Kuulkers}, and
  {Galloway}}]{Kajava14}
{Kajava}, J.~J.~E., {N{\"a}ttil{\"a}}, J., {Latvala}, O.-M., {Pursiainen}, M.,
  {Poutanen}, J., {Suleimanov}, V.~F., {Revnivtsev}, M.~G., {Kuulkers}, E.,
  {Galloway}, D.~K., Dec. 2014. {The influence of accretion geometry on the
  spectral evolution during thermonuclear (type I) X-ray bursts}. \mnras 445,
  4218--4234.

\bibitem[{{Kaptein} et~al.(2000){Kaptein}, {in't Zand}, {Kuulkers}, {Verbunt},
  {Heise}, and {Cornelisse}}]{Kaptein00}
{Kaptein}, R.~G., {in't Zand}, J.~J.~M., {Kuulkers}, E., {Verbunt}, F.,
  {Heise}, J., {Cornelisse}, R., Jun. 2000. {Discovery of 1RXS J171824.2-402934
  as an X-ray burster}. \aap 358, L71--L74.

\bibitem[{{Kuulkers} et~al.(2003){Kuulkers}, {den Hartog}, {in't Zand},
  {Verbunt}, {Harris}, and {Cocchi}}]{Kuulkers03}
{Kuulkers}, E., {den Hartog}, P.~R., {in't Zand}, J.~J.~M., {Verbunt},
  F.~W.~M., {Harris}, W.~E., {Cocchi}, M., Feb. 2003. {Photospheric radius
  expansion X-ray bursts as standard candles}. \aap 399, 663--680.

\bibitem[{{Lamb} and {Lamb}(1978)}]{Lamb78}
{Lamb}, D.~Q., {Lamb}, F.~K., Feb. 1978. {Nuclear burning in accreting neutron
  stars and X-ray bursts}. \apj 220, 291--302.

\bibitem[{{Lewin} et~al.(1993){Lewin}, {van Paradijs}, and {Taam}}]{Lewin93}
{Lewin}, W.~H.~G., {van Paradijs}, J., {Taam}, R.~E., Sep. 1993. {X-Ray
  Bursts}. \ssr 62, 223--389.

\bibitem[{{Lewin} et~al.(1995){Lewin}, {van Paradijs}, and {Taam}}]{Lewin95}
{Lewin}, W.~H.~G., {van Paradijs}, J., {Taam}, R.~E., 1995. {X-ray bursts.}
  X-ray Binaries, 175--232.

\bibitem[{{Liu} et~al.(2007){Liu}, {van Paradijs}, and {van den
  Heuvel}}]{Liu07}
{Liu}, Q.~Z., {van Paradijs}, J., {van den Heuvel}, E.~P.~J., Jul. 2007. {A
  catalogue of low-mass X-ray binaries in the Galaxy, LMC, and SMC (Fourth
  edition)}. \aap 469, 807--810.

\bibitem[{{London} et~al.(1984){London}, {Howard}, and {Taam}}]{London84}
{London}, R.~A., {Howard}, W.~M., {Taam}, R.~E., Dec. 1984. {The spectra of
  X-ray bursting neutron stars}. \apjl 287, L27--L30.

\bibitem[{{London} et~al.(1986){London}, {Taam}, and {Howard}}]{London86}
{London}, R.~A., {Taam}, R.~E., {Howard}, W.~M., Jul. 1986. {Model atmospheres
  for X-ray bursting neutron stars}. \apj 306, 170--182.

\bibitem[{{Madej} et~al.(2004){Madej}, {Joss}, and
  {R{\'o}{\.z}a{\'n}ska}}]{Madej04}
{Madej}, J., {Joss}, P.~C., {R{\'o}{\.z}a{\'n}ska}, A., Feb. 2004. {Model
  Atmospheres and X-Ray Spectra of Bursting Neutron Stars: Hydrogen-Helium
  Comptonized Spectra}. \apj 602, 904--912.

\bibitem[{{Majczyna} et~al.(2005){Majczyna}, {Madej}, {Joss}, and
  {R{\'o}{\.z}a{\'n}ska}}]{Majczyna05}
{Majczyna}, A., {Madej}, J., {Joss}, P.~C., {R{\'o}{\.z}a{\'n}ska}, A., Feb.
  2005. {Model atmospheres and X-ray spectra of bursting neutron stars. II.
  Iron rich comptonized spectra}. \aap 430, 643--654.

\bibitem[{{Manzo} et~al.(1997){Manzo}, {Giarrusso}, {Santangelo}, {Ciralli},
  {Fazio}, {Piraino}, and {Segreto}}]{Manzo97}
{Manzo}, G., {Giarrusso}, S., {Santangelo}, A., {Ciralli}, F., {Fazio}, G.,
  {Piraino}, S., {Segreto}, A., Apr. 1997. {The high pressure gas scintillation
  proportional counter on-board the BeppoSAX X-ray astronomy satellite}. \aaps
  122, 341--356.

\bibitem[{{Nakamura} et~al.(1989){Nakamura}, {Dotani}, {Inoue}, {Mitsuda},
  {Tanaka}, and {Matsuoka}}]{Nakamura89}
{Nakamura}, N., {Dotani}, T., {Inoue}, H., {Mitsuda}, K., {Tanaka}, Y.,
  {Matsuoka}, M., 1989. {TENMA observation of X-ray bursts from X1608-52}.
  \pasj 41, 617--639.

\bibitem[{{Natalucci} et~al.(2004){Natalucci}, {Bazzano}, {Cocchi}, {Ubertini},
  {Cornelisse}, {Heise}, and {in 't Zand}}]{Natalucci04}
{Natalucci}, L., {Bazzano}, A., {Cocchi}, M., {Ubertini}, P., {Cornelisse}, R.,
  {Heise}, J., {in 't Zand}, J.~J.~M., Mar. 2004. {Two spectral states of the
  transient X-ray burster SAX J1747.0-2853}. \aap 416, 699--702.

\bibitem[{{Natalucci} et~al.(2000){Natalucci}, {Bazzano}, {Cocchi}, {Ubertini},
  {Heise}, {Kuulkers}, and {in't Zand}}]{Natalucci00}
{Natalucci}, L., {Bazzano}, A., {Cocchi}, M., {Ubertini}, P., {Heise}, J.,
  {Kuulkers}, E., {in't Zand}, J.~J.~M., Nov. 2000. {Broadband Observations of
  the New X-Ray Burster SAX J1747.0-2853 during the 1998 March Outburst}. \apjl
  543, L73--L76.

\bibitem[{{{\"O}zel}(2006)}]{Ozel06}
{{\"O}zel}, F., Jun. 2006. {Soft equations of state for neutron-star matter
  ruled out by EXO 0748 - 676}. \nat 441, 1115--1117.

\bibitem[{{{\"O}zel} et~al.(2009){{\"O}zel}, {G{\"u}ver}, and
  {Psaltis}}]{Ozel09}
{{\"O}zel}, F., {G{\"u}ver}, T., {Psaltis}, D., Mar. 2009. {The Mass and Radius
  of the Neutron Star in EXO 1745-248}. \apj 693, 1775--1779.

\bibitem[{{Parmar} et~al.(1997){Parmar}, {Martin}, {Bavdaz}, {Favata},
  {Kuulkers}, {Vacanti}, {Lammers}, {Peacock}, and {Taylor}}]{Parmar97}
{Parmar}, A.~N., {Martin}, D.~D.~E., {Bavdaz}, M., {Favata}, F., {Kuulkers},
  E., {Vacanti}, G., {Lammers}, U., {Peacock}, A., {Taylor}, B.~G., Apr. 1997.
  {The low-energy concentrator spectrometer on-board the BeppoSAX X-ray
  astronomy satellite}. \aaps 122, 309--326.

\bibitem[{{Poutanen} et~al.(2014){Poutanen}, {N{\"a}ttil{\"a}}, {Kajava},
  {Latvala}, {Galloway}, {Kuulkers}, and {Suleimanov}}]{Poutanen14}
{Poutanen}, J., {N{\"a}ttil{\"a}}, J., {Kajava}, J.~J.~E., {Latvala}, O.-M.,
  {Galloway}, D.~K., {Kuulkers}, E., {Suleimanov}, V.~F., Aug. 2014. {The
  effect of accretion on the measurement of neutron star mass and radius in the
  low-mass X-ray binary 4U 1608-52}. \mnras 442, 3777--3790.

\bibitem[{{Strohmayer} and {Bildsten}(2006)}]{Strohmayer06}
{Strohmayer}, T., {Bildsten}, L., Apr. 2006. {New views of thermonuclear
  bursts}. pp. 113--156.

\bibitem[{{Suleimanov} et~al.(2011b){Suleimanov}, {Poutanen}, {Revnivtsev}, and
  {Werner}}]{Suleimanov11b}
{Suleimanov}, V., {Poutanen}, J., {Revnivtsev}, M., {Werner}, K., Dec. 2011b.
  {A Neutron Star Stiff Equation of State Derived from Cooling Phases of the
  X-Ray Burster 4U 1724-307}. \apj 742, 122.

\bibitem[{{Suleimanov} et~al.(2011a){Suleimanov}, {Poutanen}, and
  {Werner}}]{Suleimanov11a}
{Suleimanov}, V., {Poutanen}, J., {Werner}, K., Mar. 2011a. {X-ray bursting
  neutron star atmosphere models: spectra and color corrections}. \aap 527,
  A139.

\bibitem[{{Suleimanov} et~al.(2012){Suleimanov}, {Poutanen}, and
  {Werner}}]{Suleimanov12}
{Suleimanov}, V., {Poutanen}, J., {Werner}, K., Sep. 2012. {X-ray bursting
  neutron star atmosphere models using an exact relativistic kinetic equation
  for Compton scattering}. \aap 545, A120.

\bibitem[{{Syunyaev} and {Titarchuk}(1986)}]{Syunyaev86}
{Syunyaev}, R.~A., {Titarchuk}, L.~G., Dec. 1986. {On the Spectra of X-Ray
  Bursters}. Soviet Astronomy Letters 12, 359--364.

\bibitem[{{Titarchuk}(1994)}]{Titarchuk94}
{Titarchuk}, L., Jul. 1994. {On the specta of X-ray bursters: Expansion and
  contraction stages}. \apj 429, 340--355.

\bibitem[{{van Paradijs}(1978)}]{vanParadijs78}
{van Paradijs}, J., Aug. 1978. {Average properties of X-ray burst sources}.
  \nat 274, 650--653.

\bibitem[{{van Straaten} et~al.(2001){van Straaten}, {van der Klis},
  {Kuulkers}, and {M{\'e}ndez}}]{Van01}
{van Straaten}, S., {van der Klis}, M., {Kuulkers}, E., {M{\'e}ndez}, M., Apr.
  2001. {An Atlas of Burst Oscillations and Spectral Properties in 4U 1728-34}.
  \apj 551, 907--920.

\end{thebibliography}

\begin{figure*}
\begin{minipage}{0.35\textwidth}
\includegraphics[height=3.in,width=3.0in,angle=0]{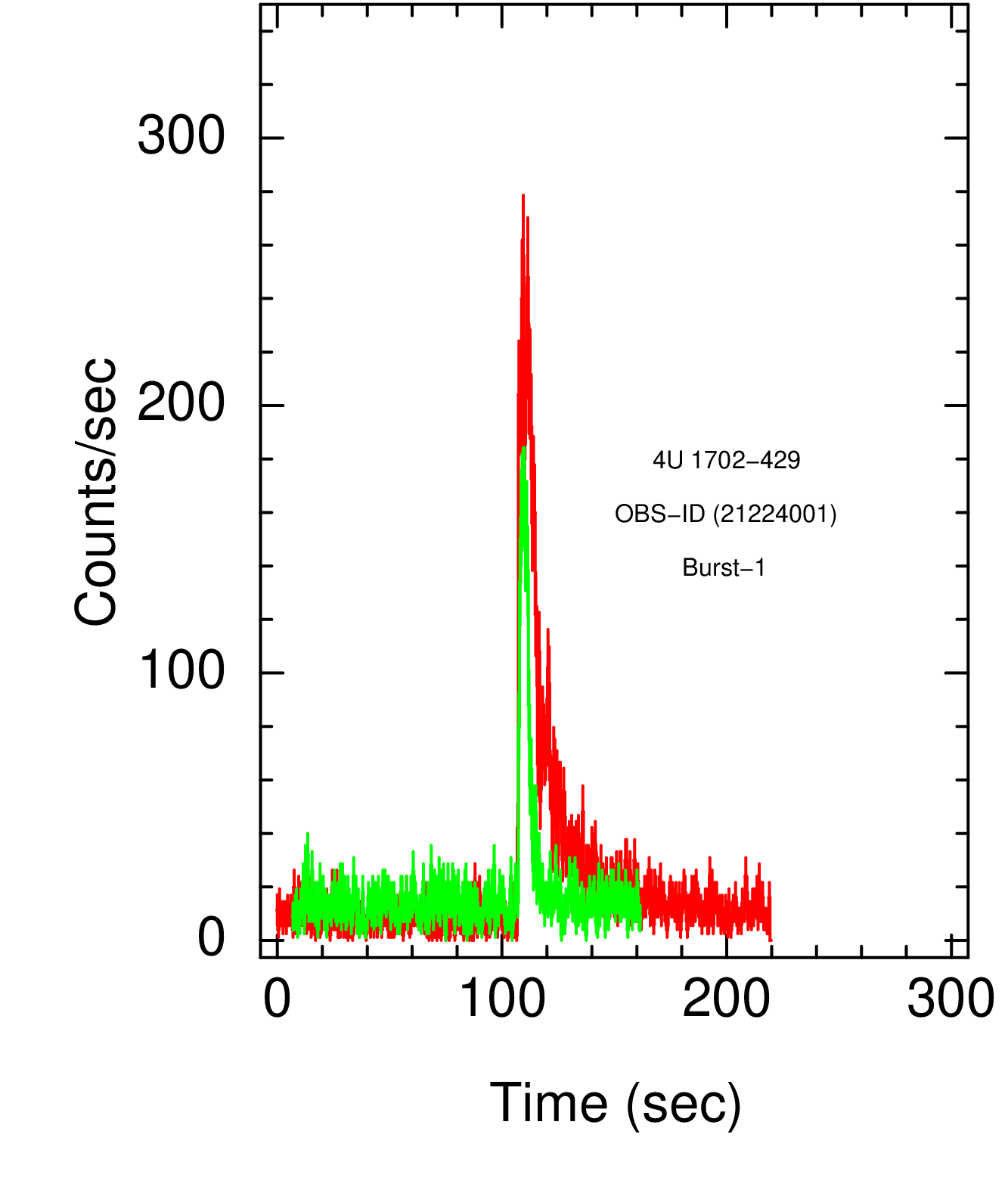}
\end{minipage}
 \hspace{0.25\linewidth}
\begin{minipage}{0.35\textwidth}
 \includegraphics[height=3.in,width=3.0in,angle=0]{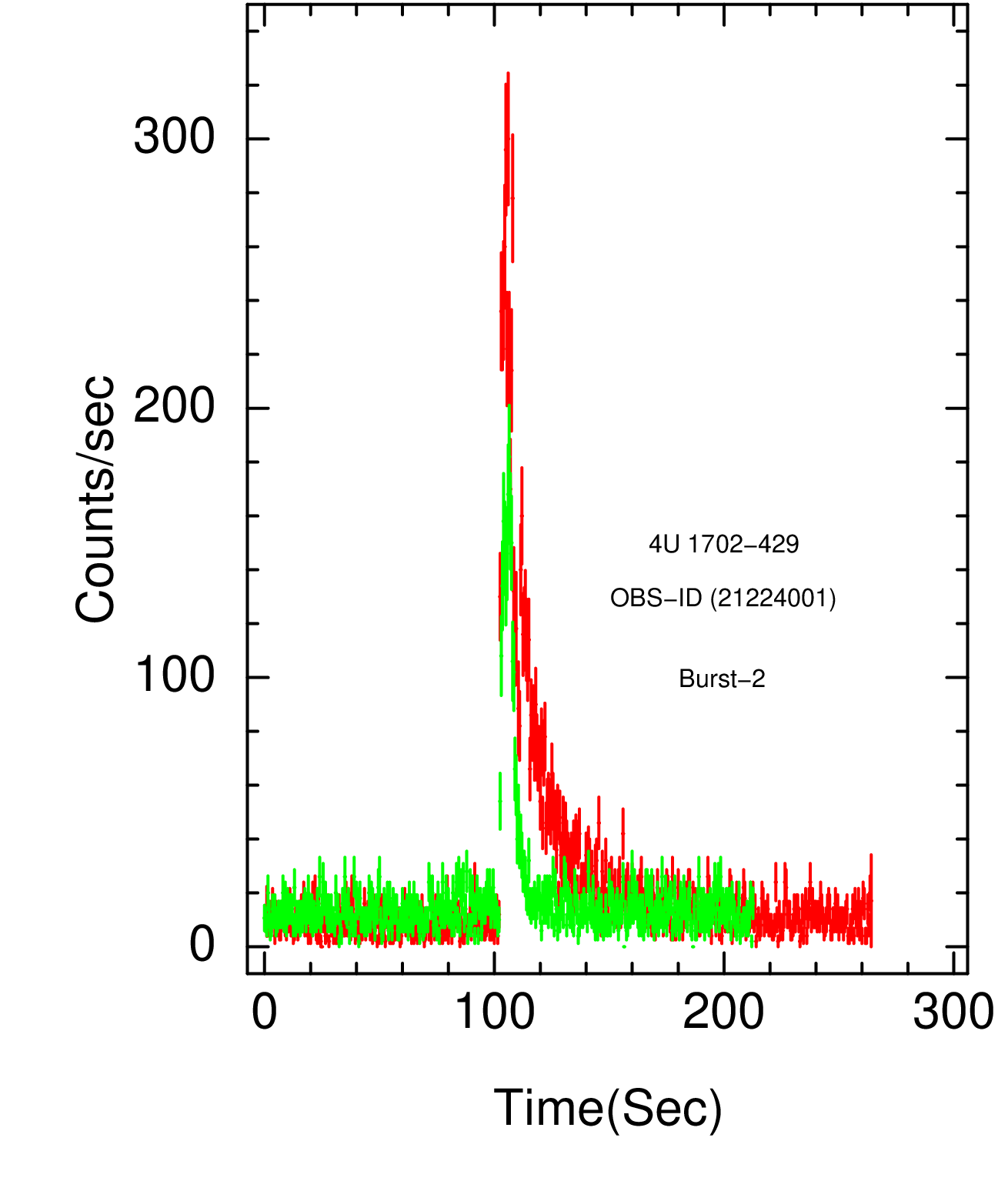}
 \end{minipage}
  \hspace{0.25\linewidth}
 \begin{minipage}{0.35\textwidth}
 \includegraphics[height=3.in,width=3.0in,angle=0]{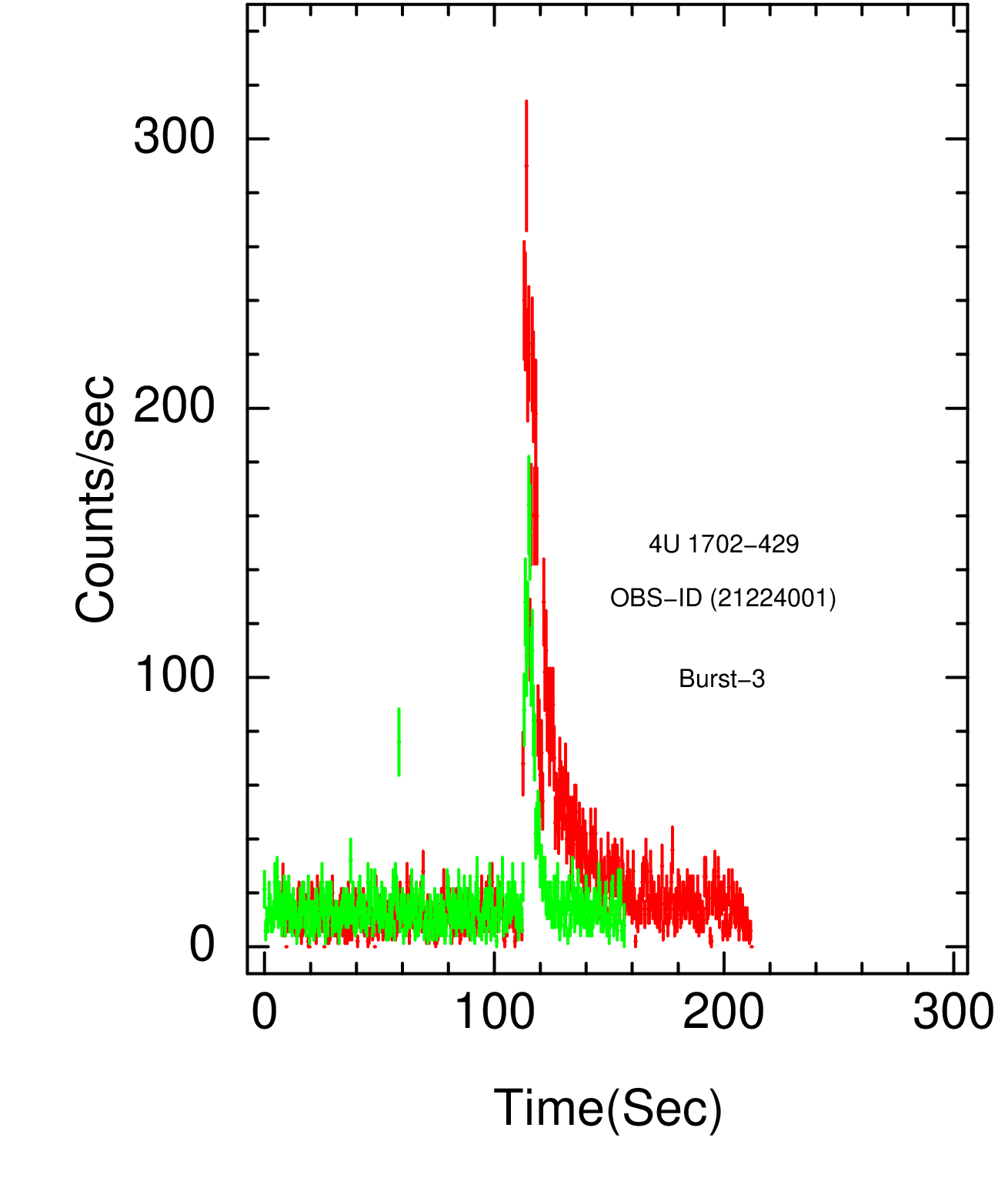}
 \end{minipage}
 \hspace{0.25\linewidth}
 \begin{minipage}{0.35\textwidth}
 \includegraphics[height=3.in,width=3.0in,angle=0]{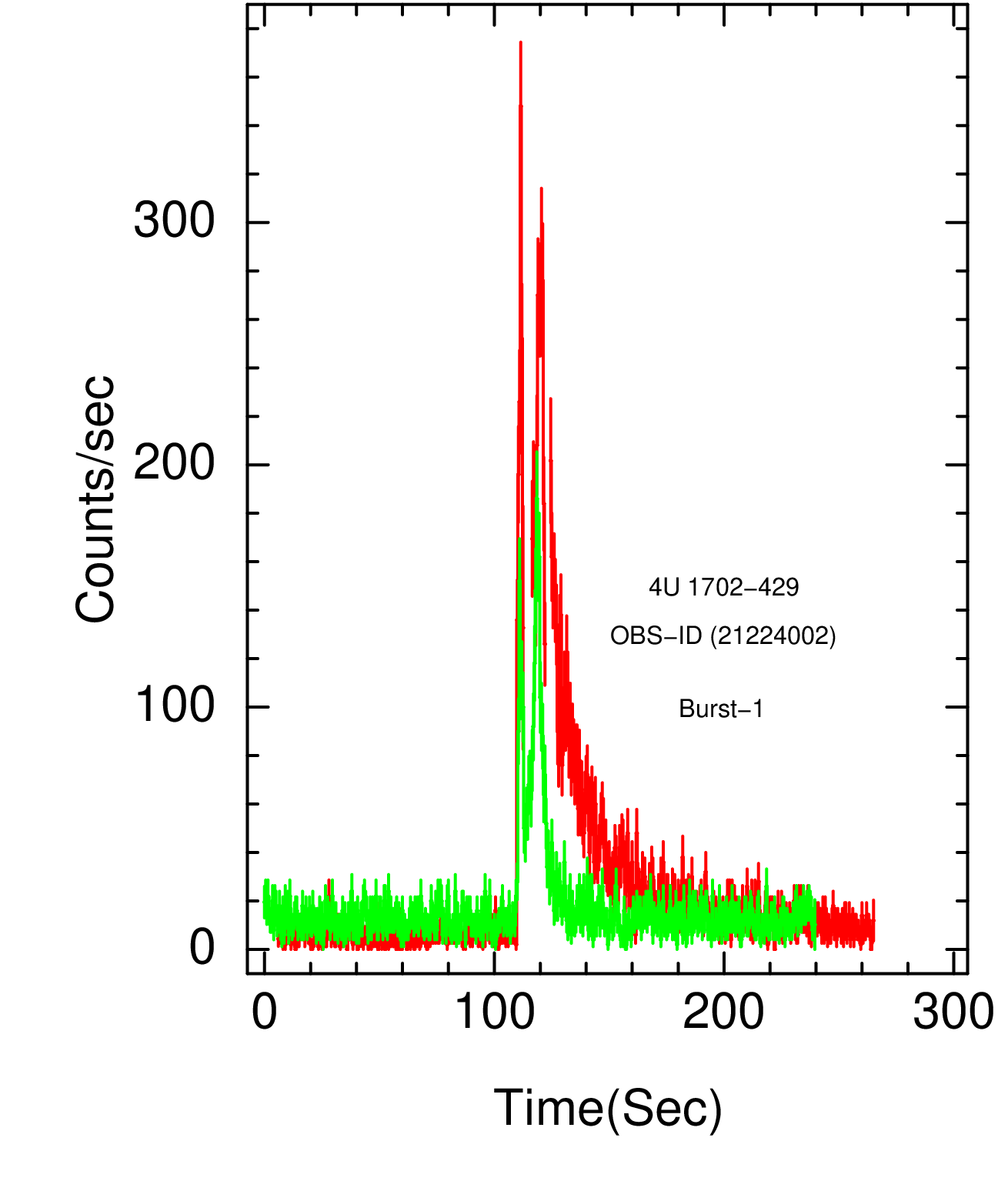}
 \end{minipage}
 \hspace{0.35\linewidth}
  \begin{minipage}{0.3\textwidth}
 \includegraphics[height=3.in,width=3.0in,angle=0]{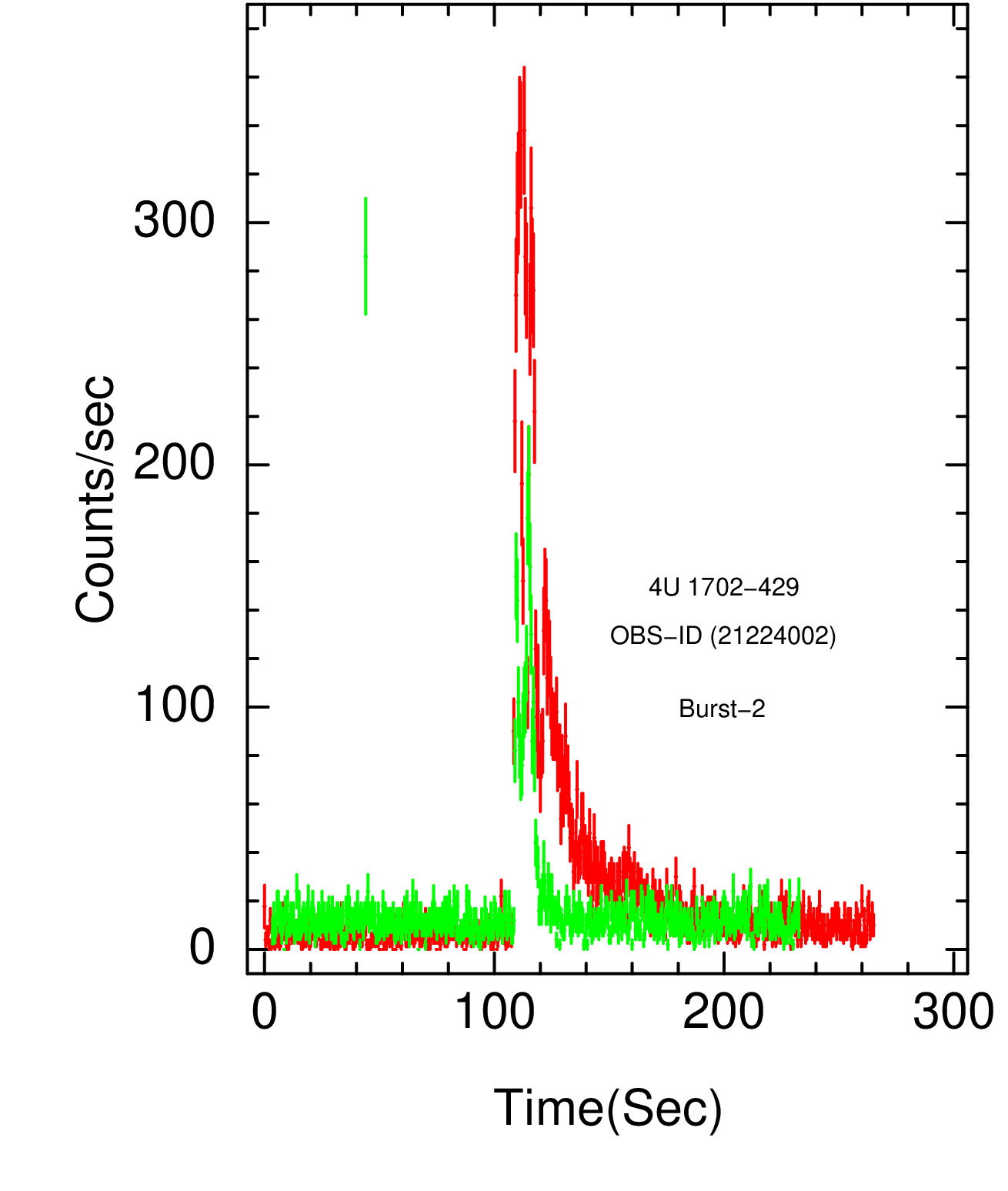}
 \end{minipage}
\hspace{0.35\linewidth}
 \begin{minipage}{0.3\textwidth}
 \includegraphics[height=3.in,width=3.0in,angle=0]{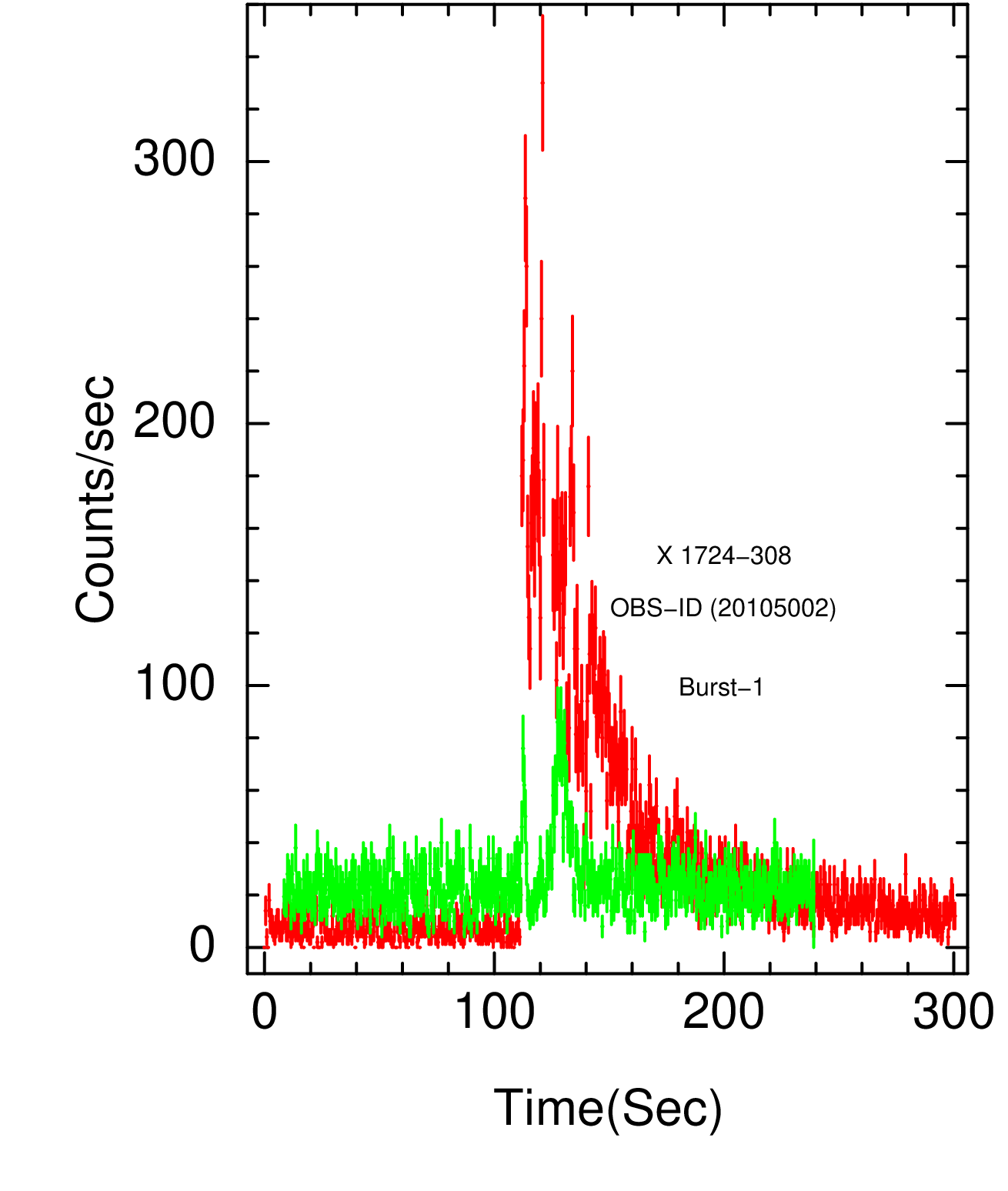}
 \end{minipage}
  \label{lcurves}
\end{figure*} 
 
\begin{figure*}
\begin{minipage}{0.35\textwidth}
 \includegraphics[height=3.in,width=3.0in,angle=0]{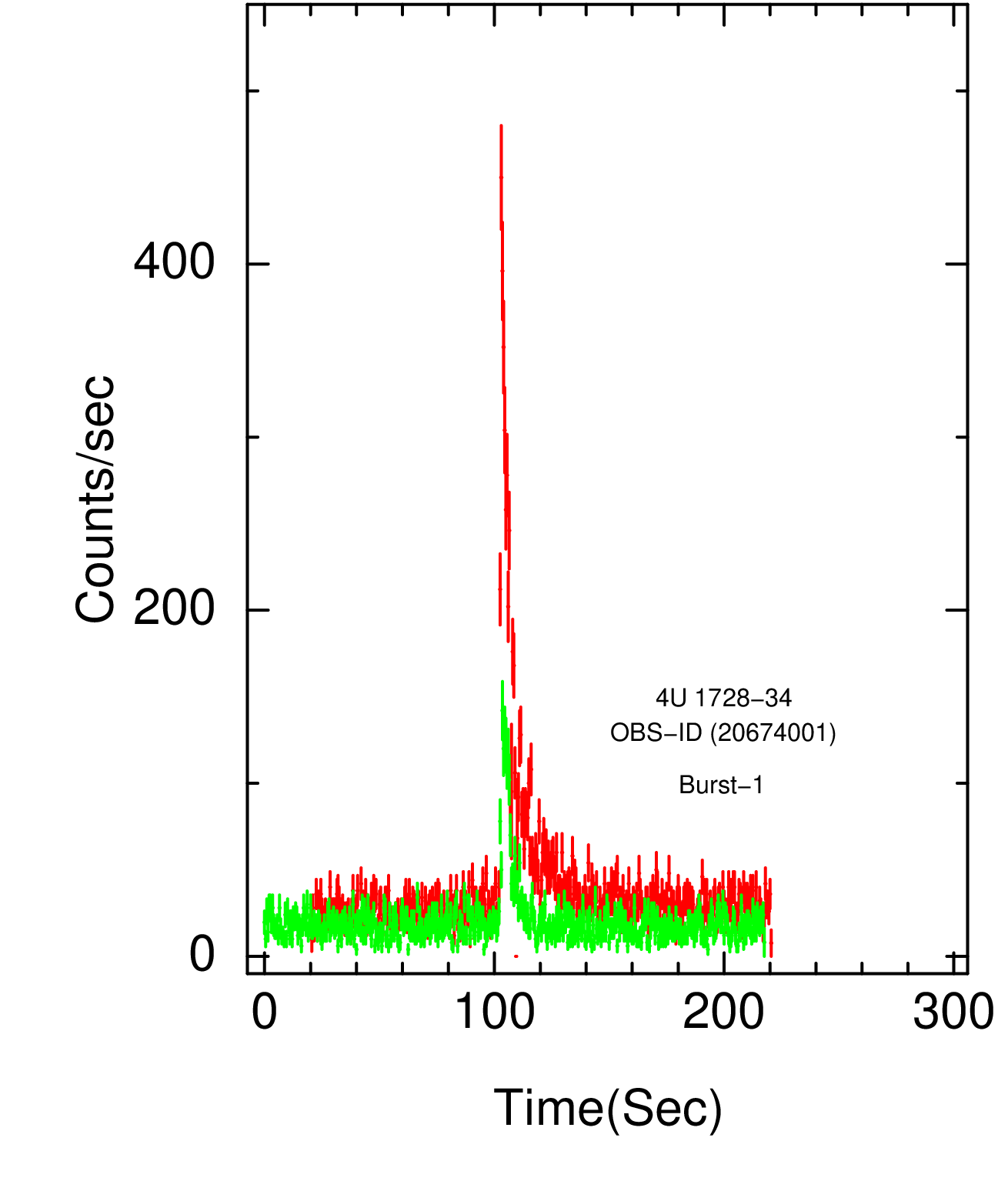}
 \end{minipage}
\hspace{0.25\textwidth}
 \begin{minipage}{0.35\textwidth}
 \includegraphics[height=3.in,width=3.0in,angle=0]{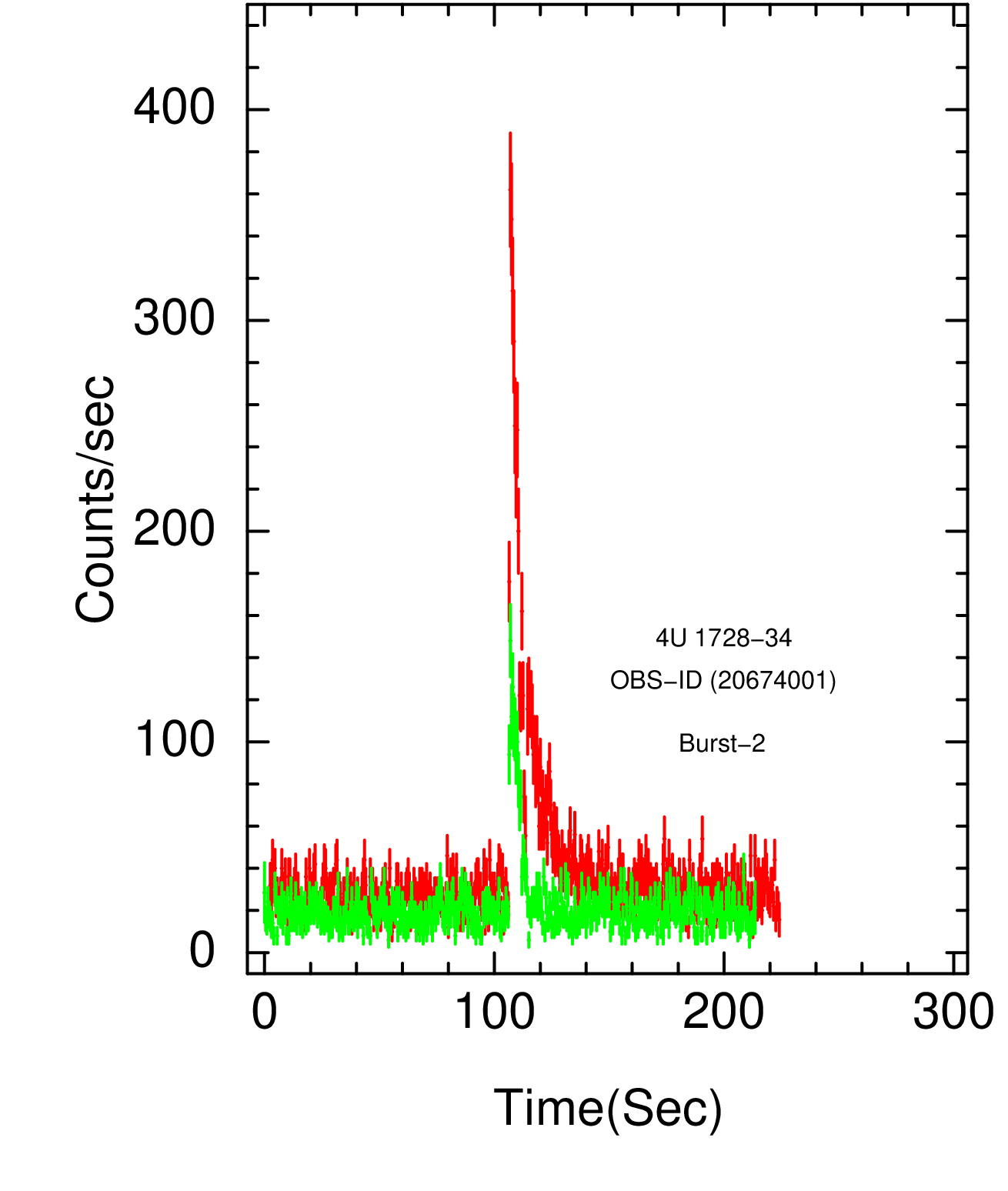}
 \end{minipage}
 \hspace{0.25\textwidth}
  \begin{minipage}{0.35\textwidth}
 \includegraphics[height=3.in,width=3.0in,angle=0]{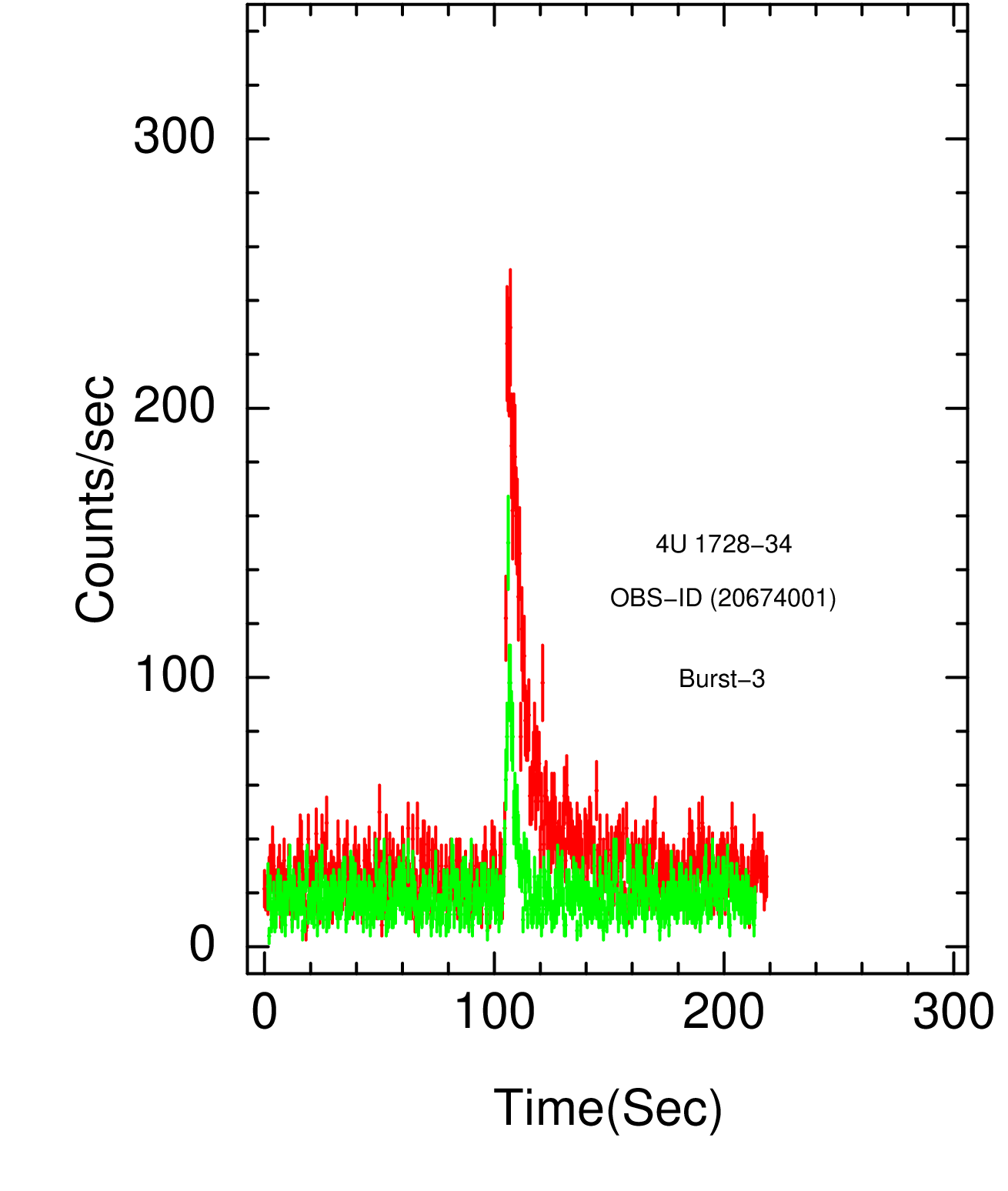}
 \end{minipage}
  \hspace{0.25\textwidth}
 \begin{minipage}{0.35\textwidth}
 \includegraphics[height=3.in,width=3.0in,angle=0]{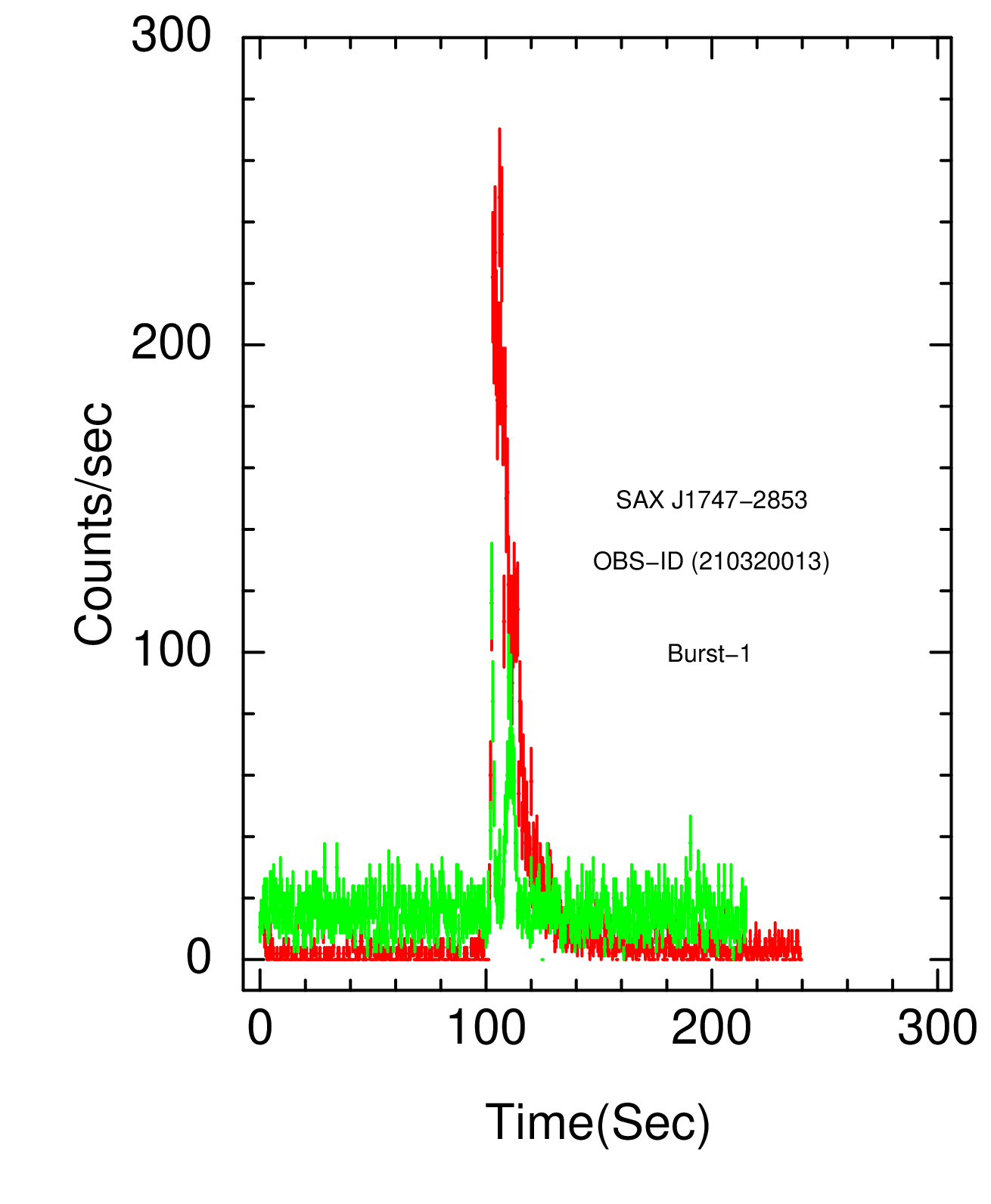}
 \end{minipage}
  \hspace{0.25\textwidth}
  \begin{minipage}{0.35\textwidth}
 \includegraphics[height=3.in,width=3.0in,angle=0]{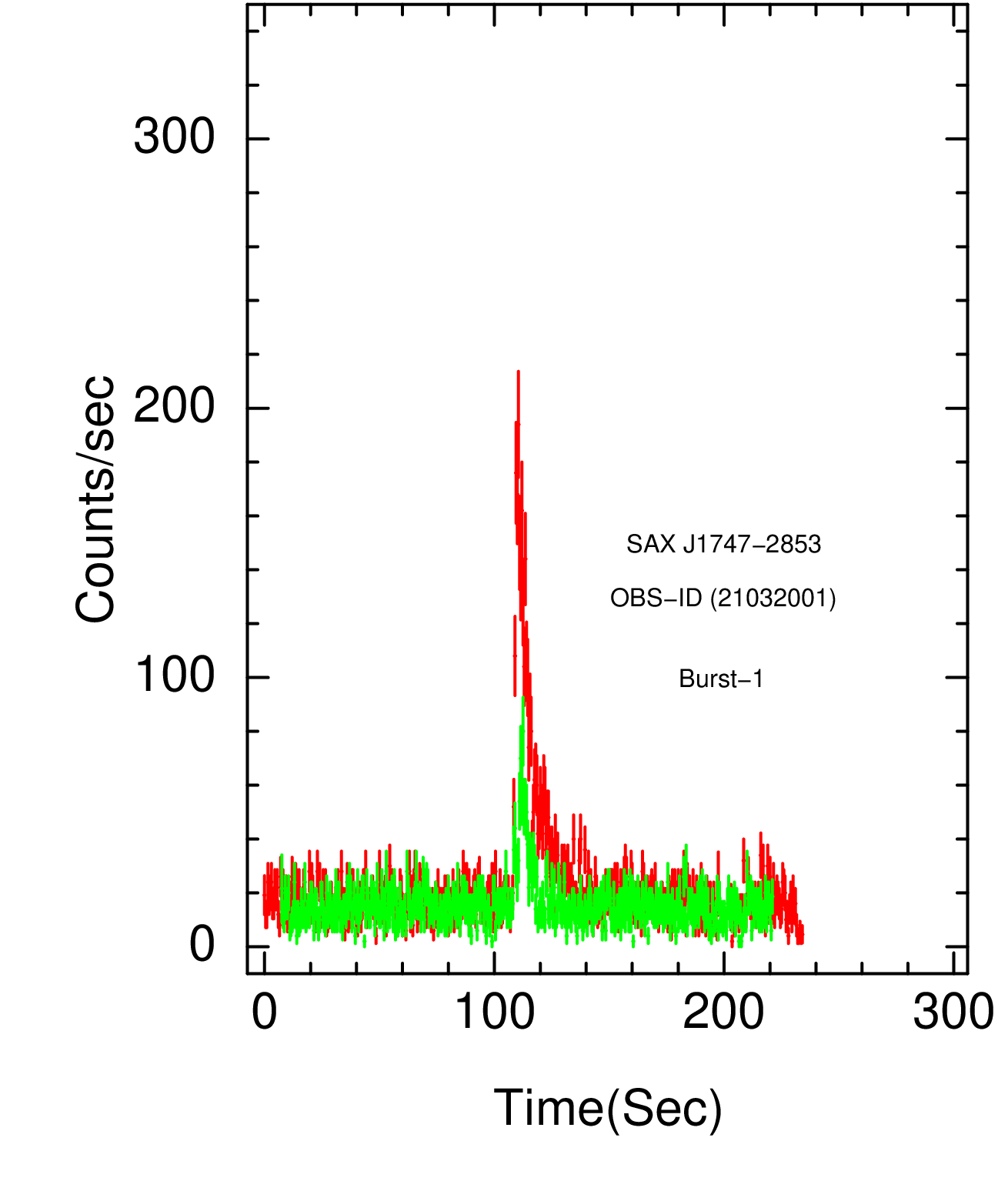}
 \end{minipage}
 \hspace{0.25\textwidth}
 \begin{minipage}{0.35\textwidth}
  \includegraphics[height=3.in,width=3.0in,angle=0]{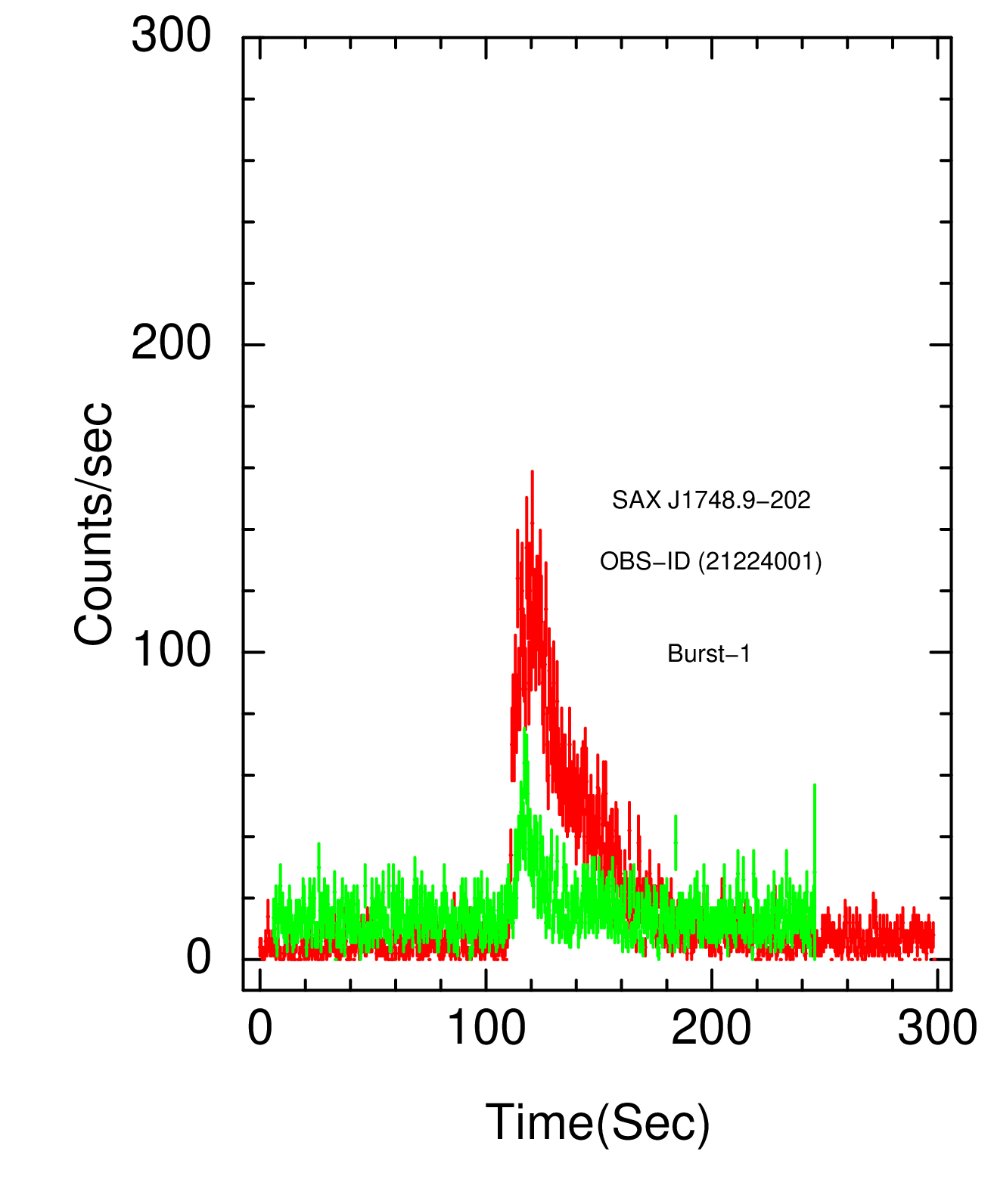}
 \end{minipage} 
 \label{lcurves}
\end{figure*}

\begin{figure*}
 \begin{minipage}{0.35\textwidth} 
\includegraphics[height=3.in,width=3.0in,angle=0]{fig-1m.ps}
\end{minipage}
 \hspace{0.25\textwidth}
 \begin{minipage}{0.35\textwidth}
 \includegraphics[height=3.in,width=3.0in,angle=0]{fig-1n.ps}
 \end{minipage}
 \hspace{0.25\textwidth}
  \begin{minipage}{0.35\textwidth}
 \includegraphics[height=3.in,width=3.0in,angle=0]{fig-1o.ps}
 \end{minipage}
\hspace{0.25\textwidth}
 \begin{minipage}{0.35\textwidth}
 \includegraphics[height=3.in,width=3.0in,angle=0]{fig-1p.ps}
 \end{minipage}
 \hspace{0.25\textwidth}
 \begin{minipage}{0.35\textwidth}
 \includegraphics[height=3.in,width=3.0in,angle=0]{fig-1q.ps}
 \end{minipage}
\hspace{0.25\textwidth}
\begin{minipage}{0.35\textwidth}
 \includegraphics[height=3.in,width=3.0in,angle=0]{fig-1r.ps}
 \end{minipage}

\label{lcurves}
\end{figure*} 

\begin{figure*}

   \begin{minipage}{0.35\textwidth}
 \includegraphics[height=3.in,width=3.0in,angle=0]{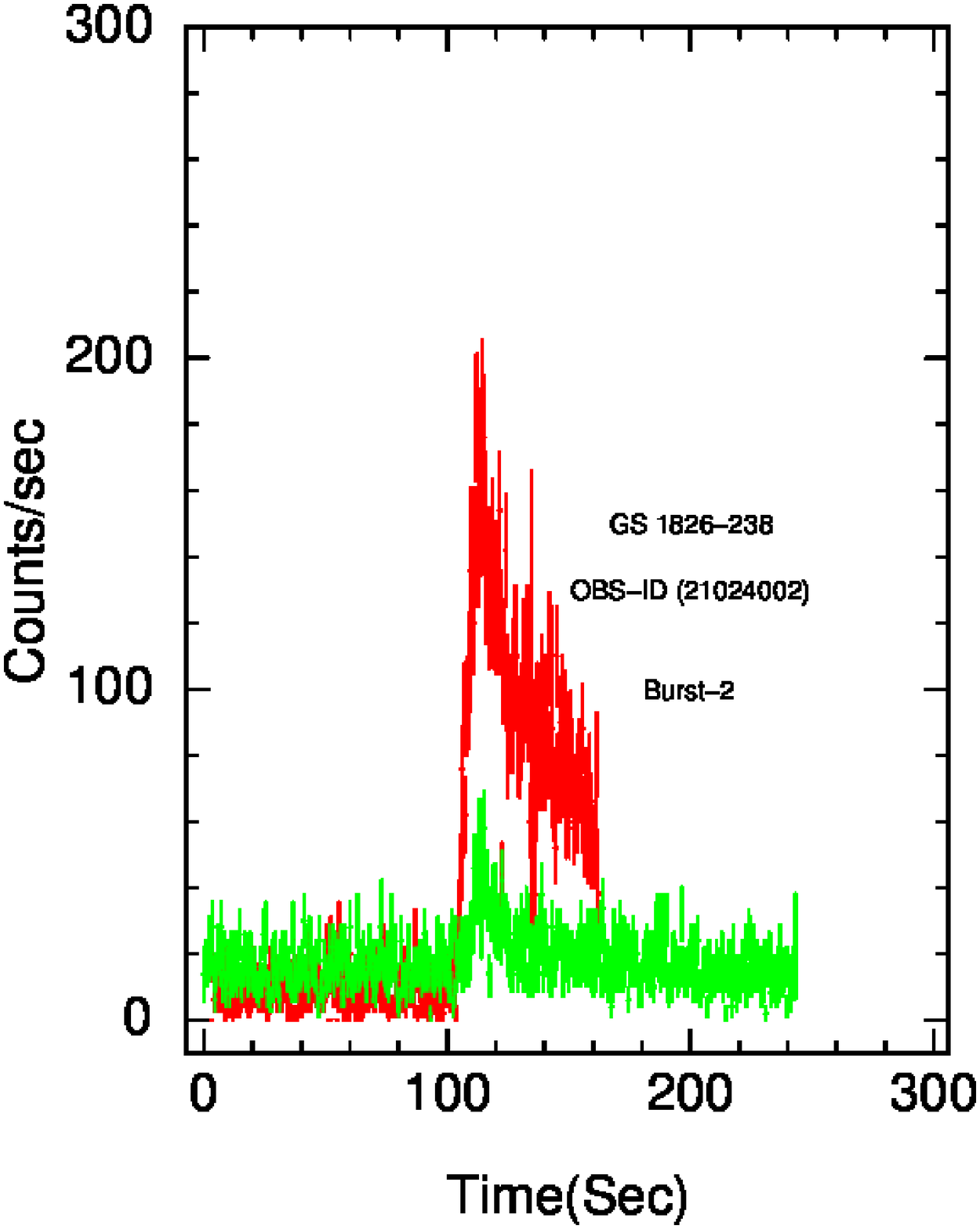}
 \end{minipage}
 \hspace{0.25\textwidth}
 \begin{minipage}{0.35\textwidth}
 \includegraphics[height=3.in,width=3.0in,angle=0]{fig-1t.ps}
 \end{minipage}
 \hspace{0.3\textwidth}
 \begin{minipage}{0.35\textwidth}
 \includegraphics[height=3.in,width=3.0in,angle=0]{fig-1u.ps}
 \end{minipage}
 \hspace{0.3\textwidth}
 \begin{minipage}{0.35\textwidth}
 \includegraphics[height=3.in,width=3.0in,angle=0]{fig-1v.ps}
 \end{minipage}
\caption{Lightcurves during bursts with 100 seconds before and after the burst,
created using data from \emph{MECS} and \emph{PDS} with binsize=0.5 seconds,
red color curves correspond to \emph{MECS} while green is for \emph{PDS} bursts.}
\label{lcurves}
\end{figure*}

\begin{figure*}
\begin{minipage}{0.35\textwidth}
\includegraphics[height=3.in,width=3.0in,angle=0]{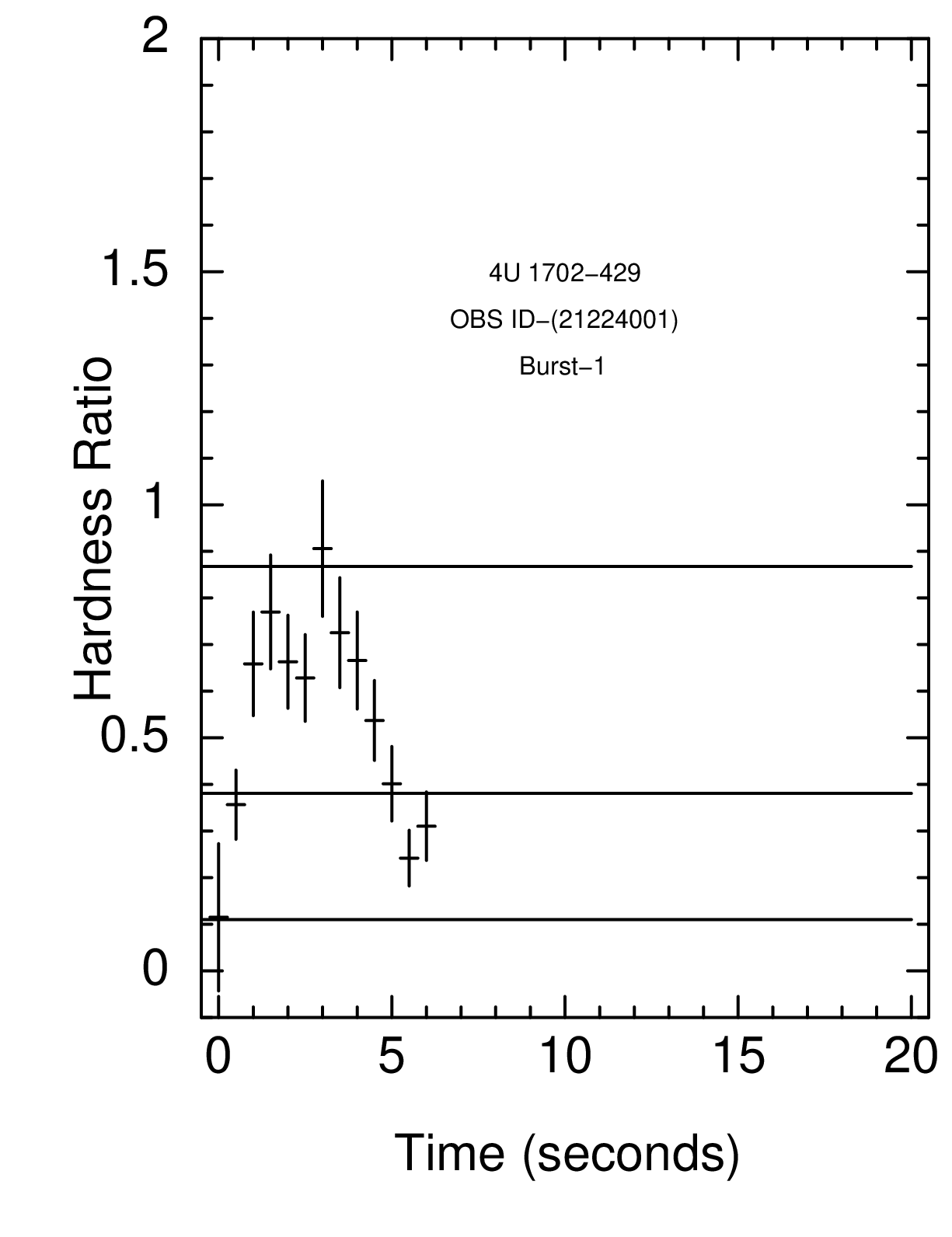}
 \end{minipage}
 \hspace{0.2\linewidth}
 \begin{minipage}{0.35\textwidth}
 \includegraphics[height=3.in,width=3.0in,angle=0]{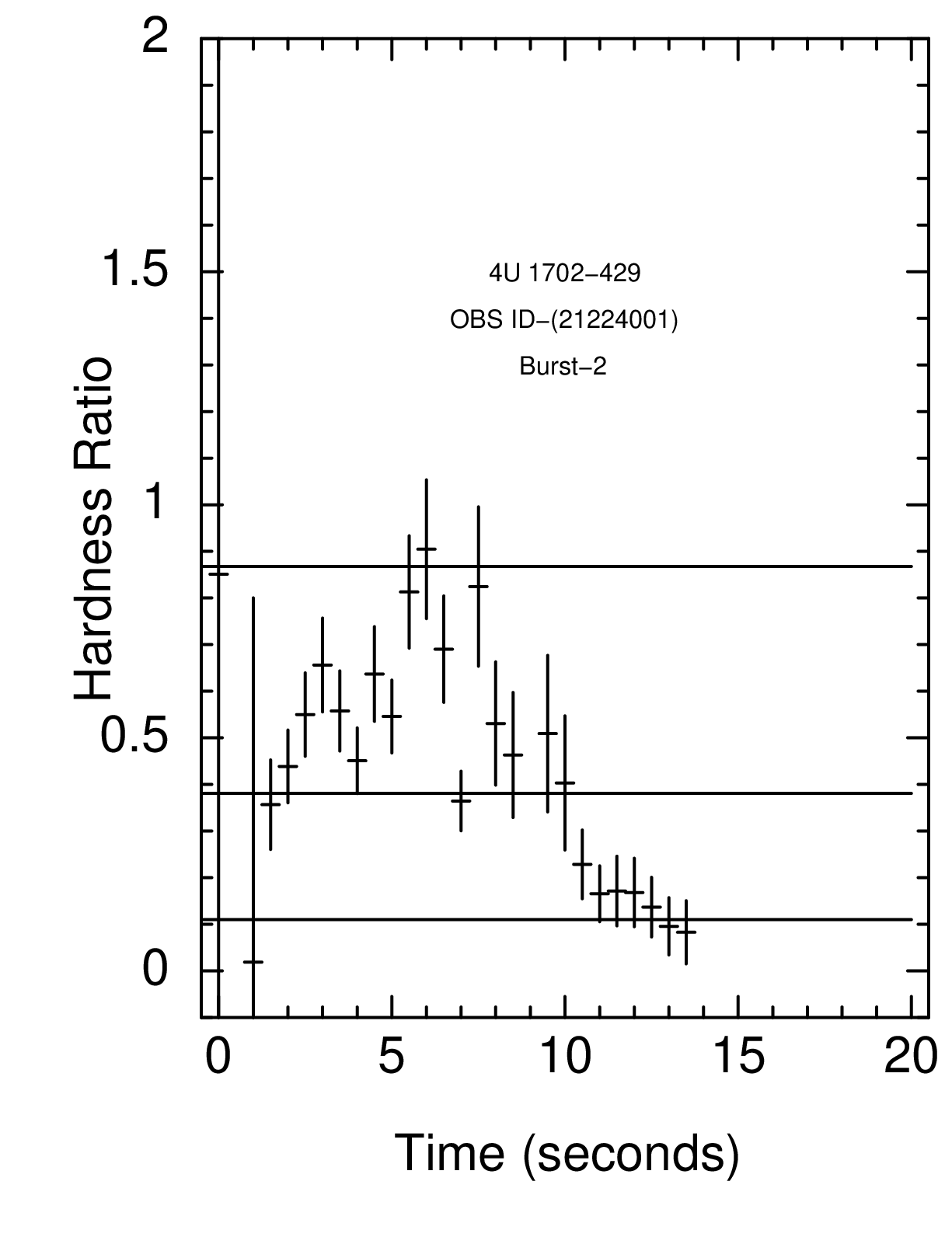}
 \end{minipage}
 \hspace{0.2\linewidth}
\begin{minipage}{0.43\textwidth}
 \includegraphics[height=3.in,width=3.0in,angle=0]{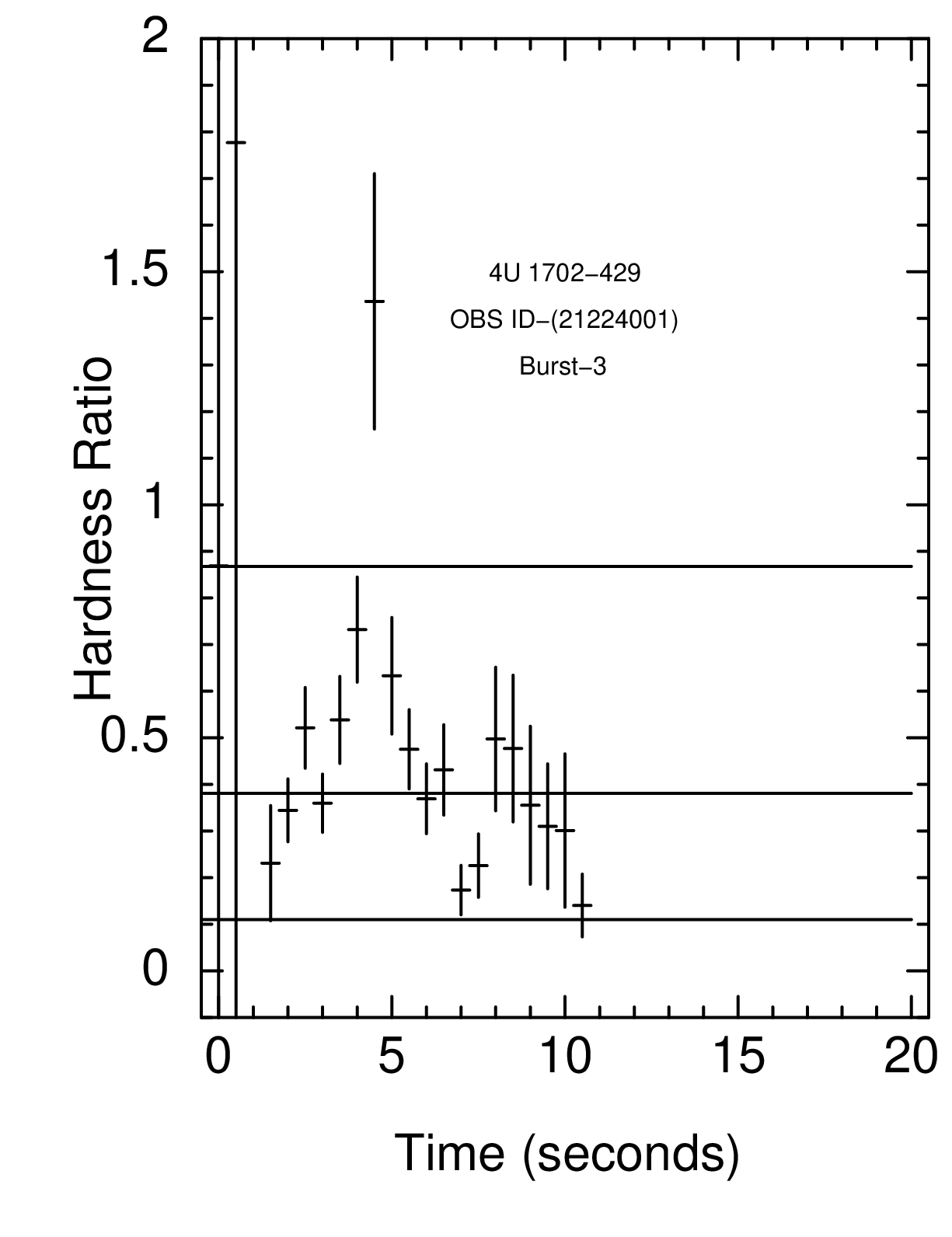}
 \end{minipage}
\hspace{0.2\linewidth}
\begin{minipage}{0.43\textwidth}
\includegraphics[height=3.in,width=3.0in,angle=0]{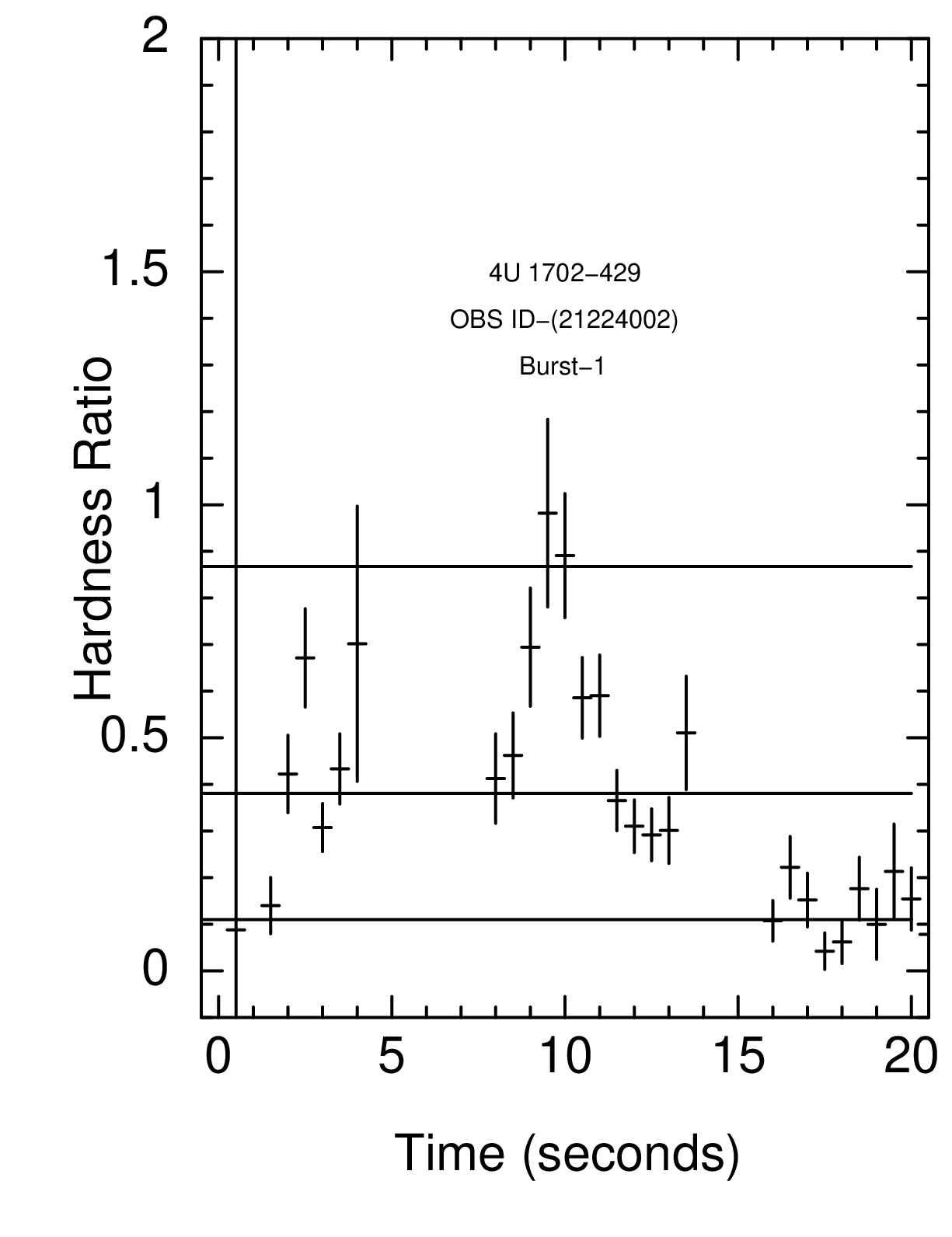}
\end{minipage}
\hspace{0.2\linewidth}
\begin{minipage}{0.35\textwidth}
\includegraphics[height=3.in,width=3.0in,angle=0]{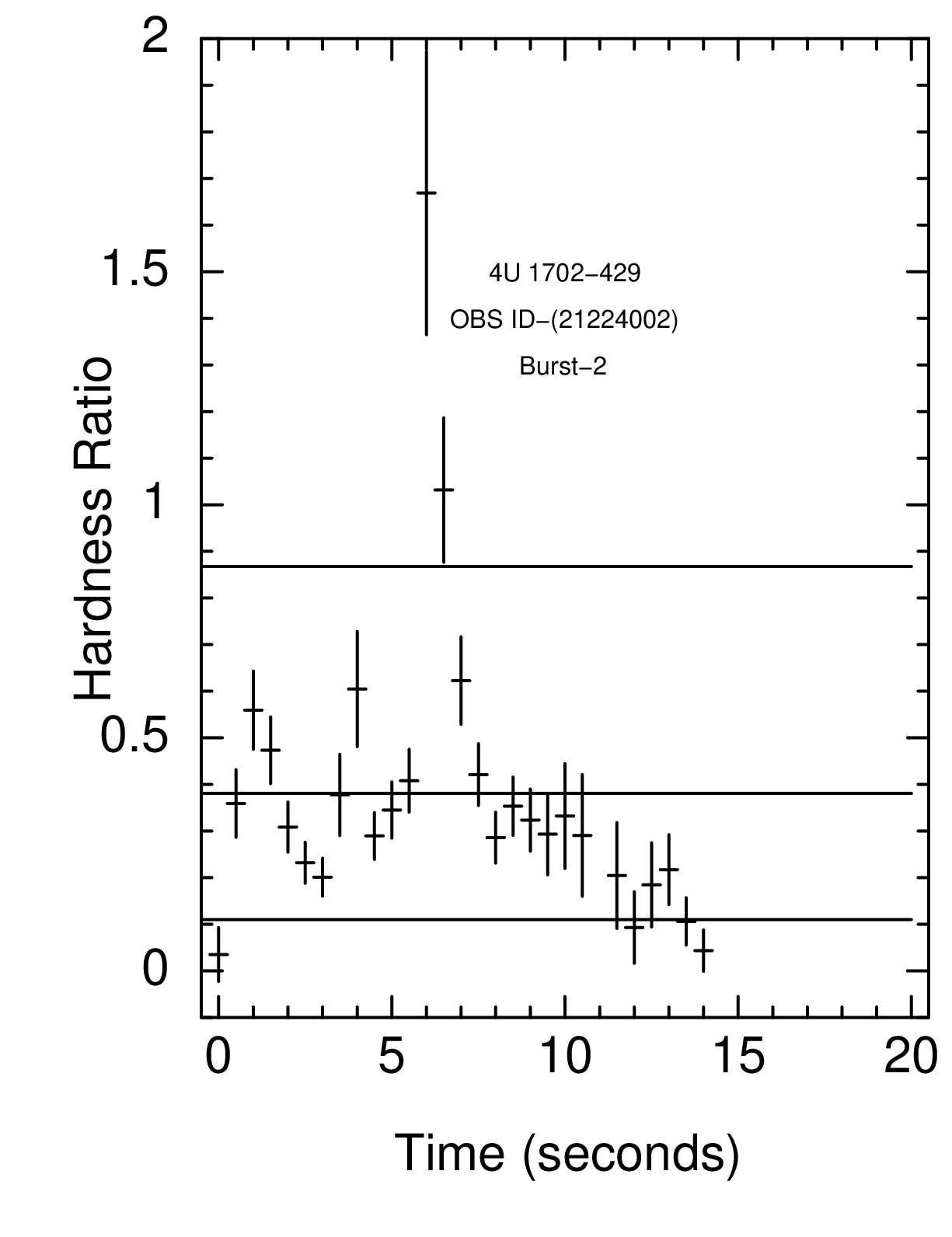}
\end{minipage}
\hspace{0.3\linewidth}
\begin{minipage}{0.2\textwidth}
\includegraphics[height=3.in,width=3.0in,angle=0]{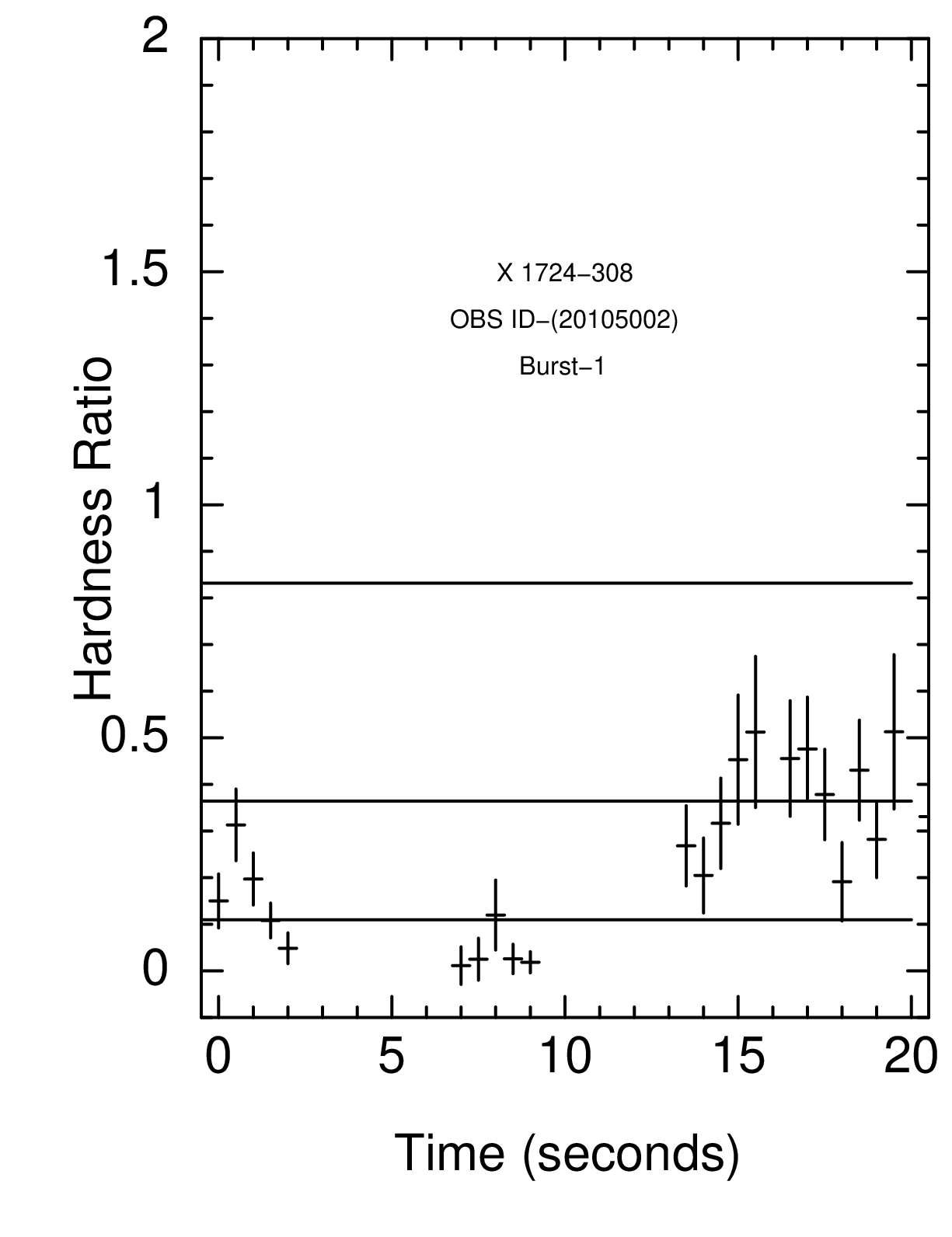}
\end{minipage}
\label{HR}
 \end{figure*}
 
\begin{figure*}
\begin{minipage}{0.43\textwidth}
\includegraphics[height=3.in,width=3.0in,angle=0]{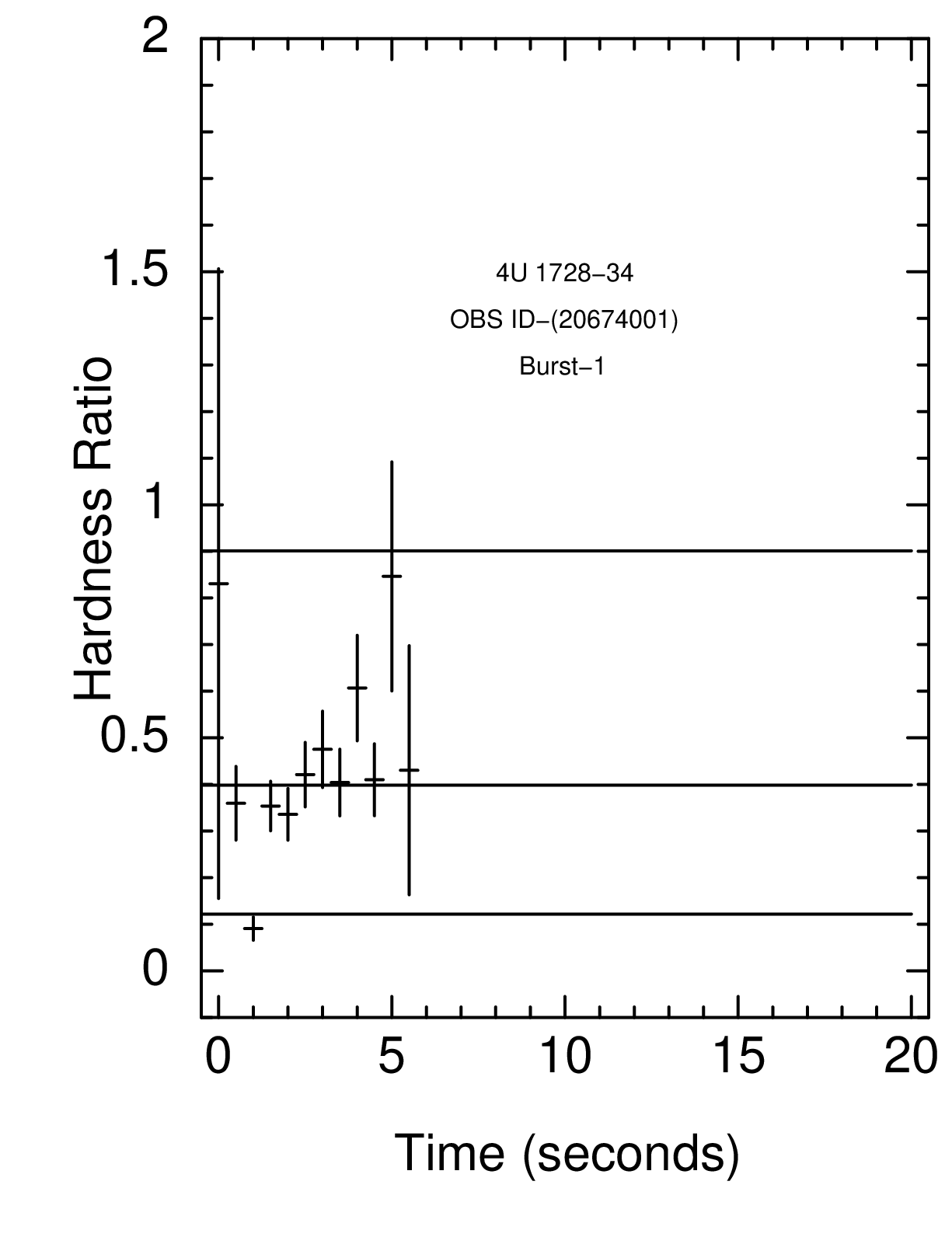}
 \end{minipage}
 \hspace{0.2\textwidth}
\begin{minipage}{0.43\textwidth}
\includegraphics[height=3.in,width=3.0in,angle=0]{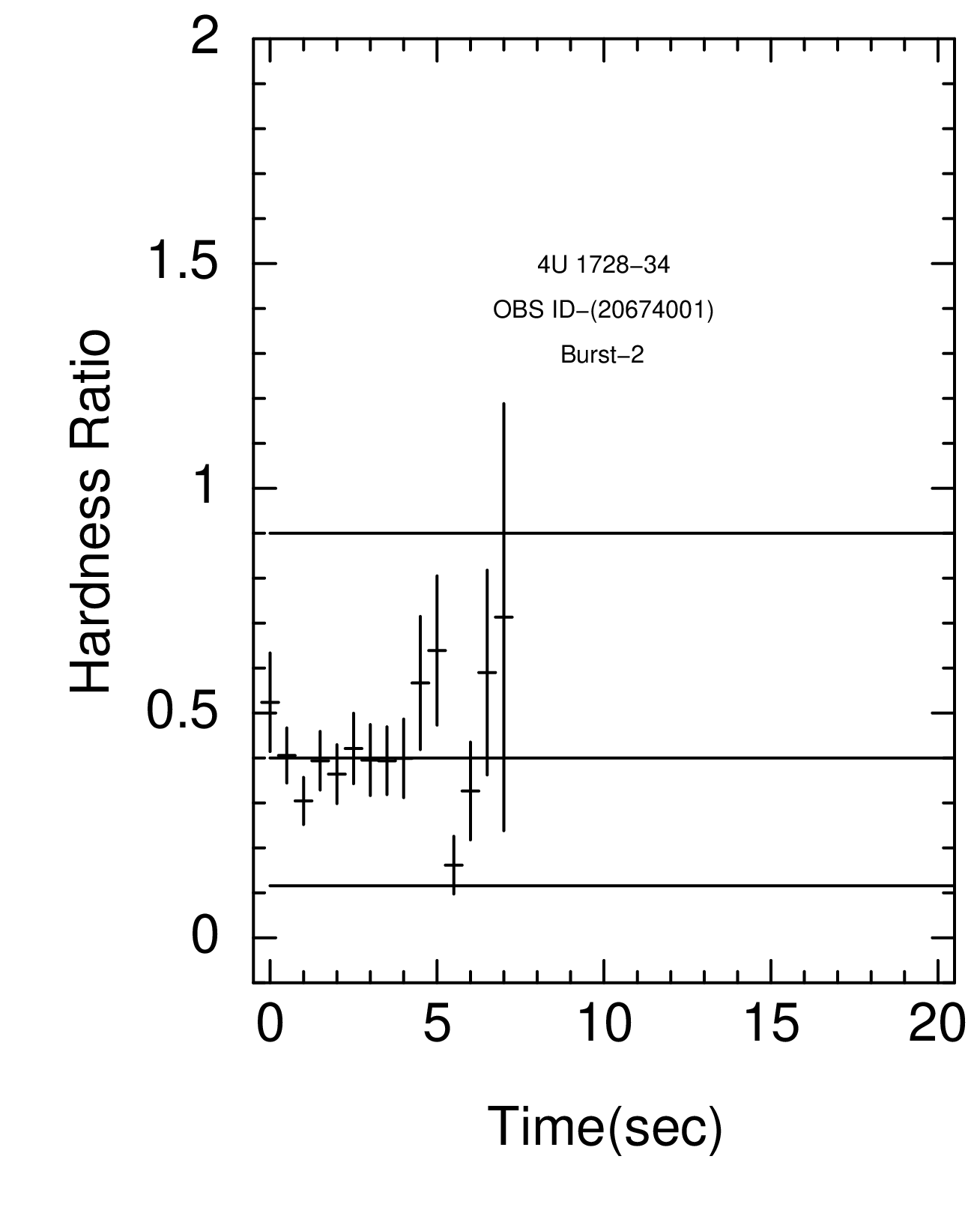}
\end{minipage}
 \hspace{0.2\textwidth}
 \begin{minipage}{0.35\textwidth}
\includegraphics[height=3.in,width=3.0in,angle=0]{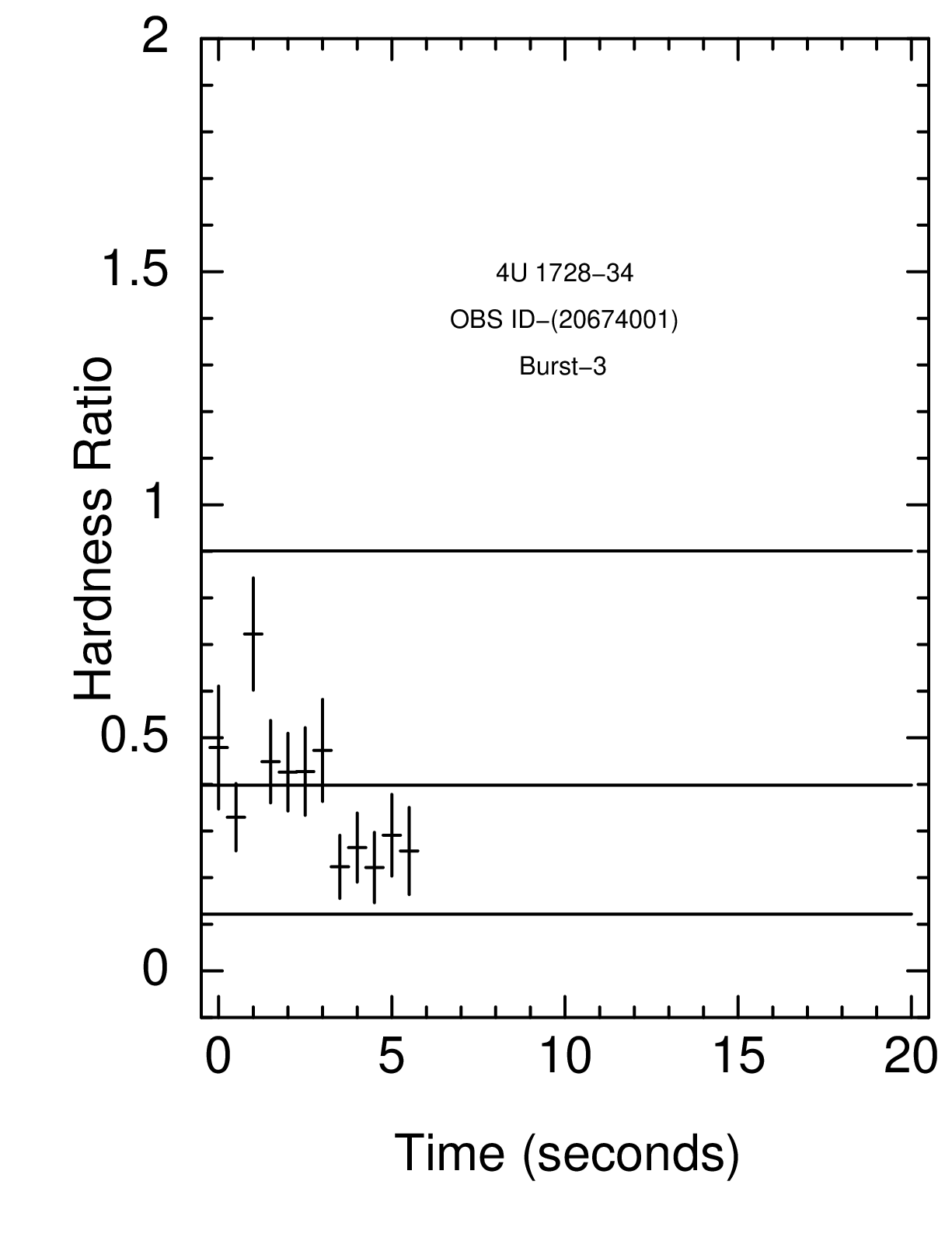}
\end{minipage}
\hspace{0.2\textwidth}
\begin{minipage}{0.35\textwidth}
\includegraphics[height=3.in,width=3.0in,angle=0]{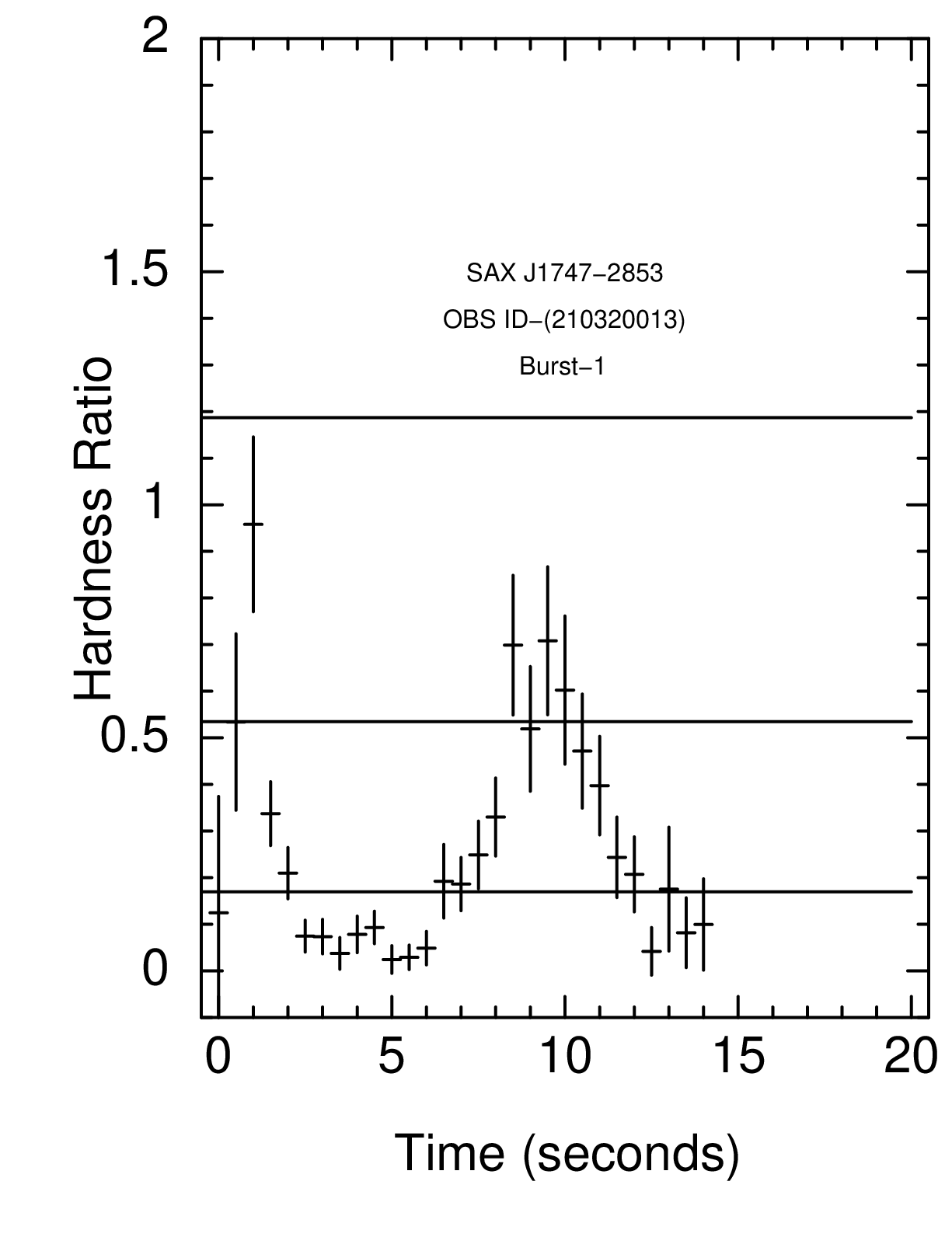}
\end{minipage}
\hspace{0.2\textwidth}
\begin{minipage}{0.43\textwidth}
\includegraphics[height=3.in,width=3.0in,angle=0]{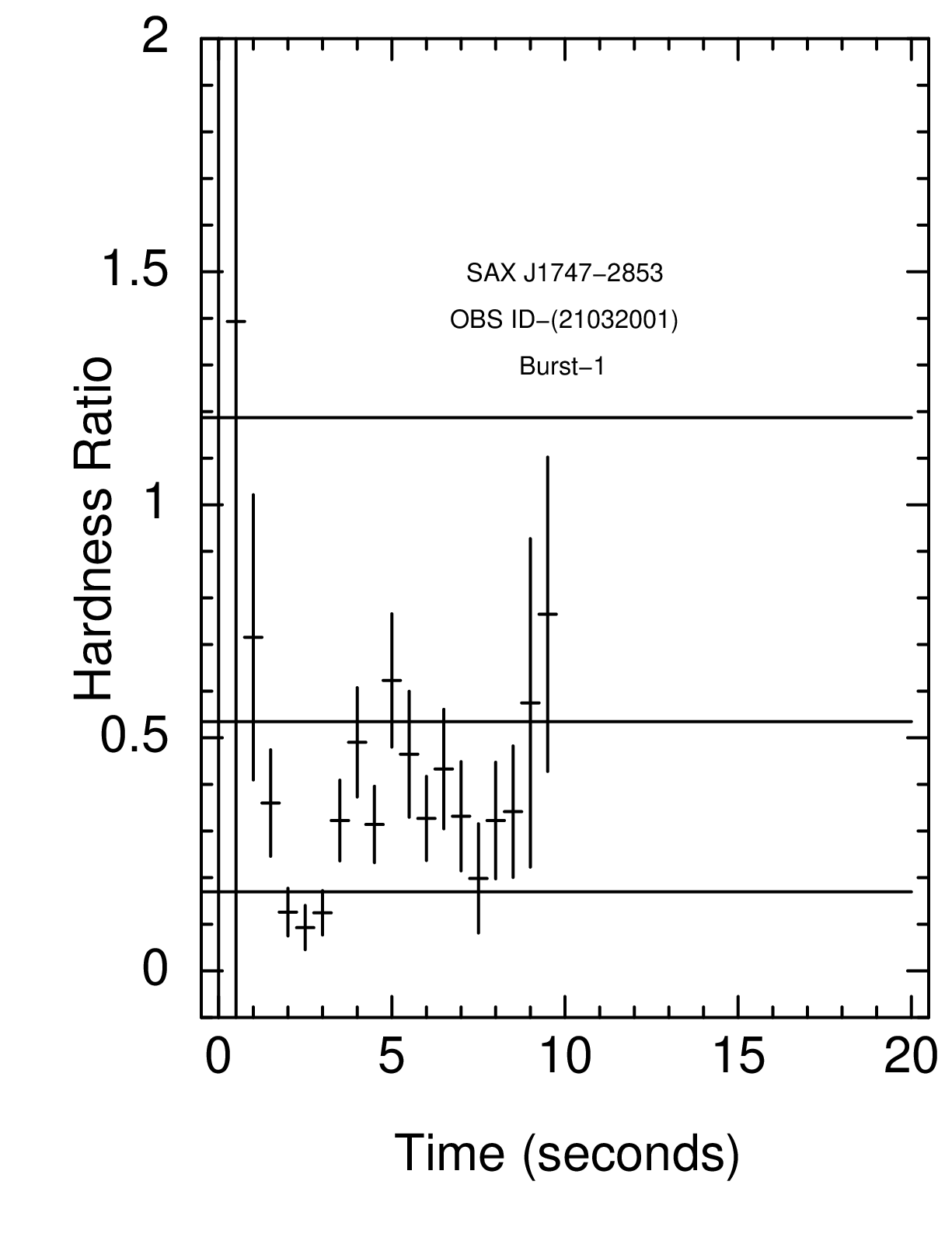}
\end{minipage}
\hspace{0.2\textwidth}
\begin{minipage}{0.43\textwidth}
\includegraphics[height=3.in,width=3.0in,angle=0]{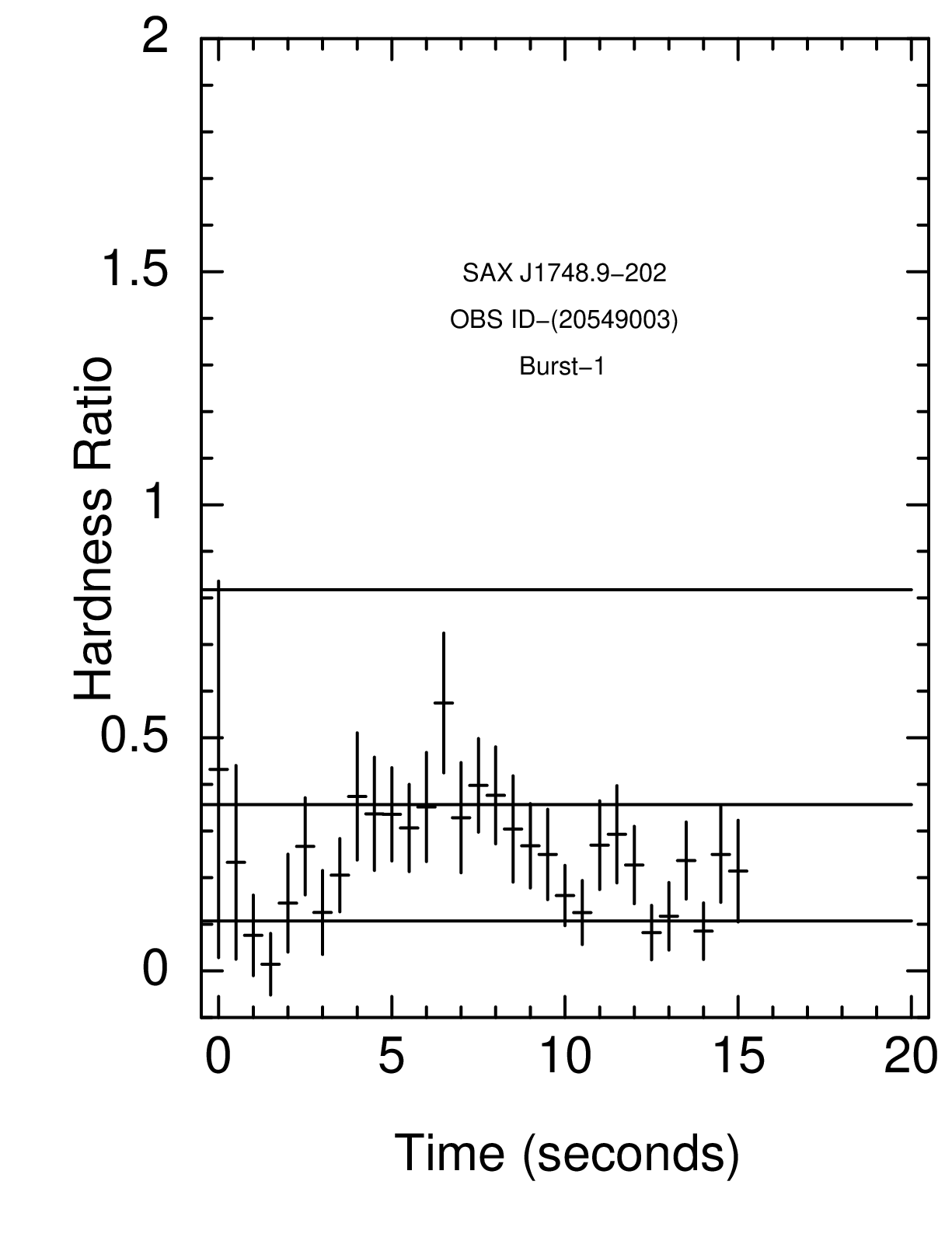}
\end{minipage}

 \label{HR}
 \end{figure*}

\begin{figure*}
\begin{minipage}{0.35\textwidth}
\includegraphics[height=3.in,width=3.0in,angle=0]{fig-2m.ps}
\end{minipage}
\hspace{0.2\textwidth}
\begin{minipage}{0.35\textwidth}
\includegraphics[height=3.in,width=3.0in,angle=0]{fig-2n.ps}
\end{minipage}
\hspace{0.2\textwidth}
\begin{minipage}{0.43\textwidth}
\includegraphics[height=3.in,width=3.0in,angle=0]{fig-2o.ps}
\end{minipage}
\hspace{0.2\textwidth}
\begin{minipage}{0.43\textwidth}
\includegraphics[height=3.in,width=3.0in,angle=0]{fig-2p.ps}
\end{minipage}
\hspace{0.2\textwidth}
\begin{minipage}{0.35\textwidth}
\includegraphics[height=3.0in,width=3.0in,angle=0]{fig-2q.ps}
\end{minipage}
\hspace{0.3\textwidth}
\begin{minipage}{0.35\textwidth}
\includegraphics[height=3.0in,width=3.0in,angle=0]{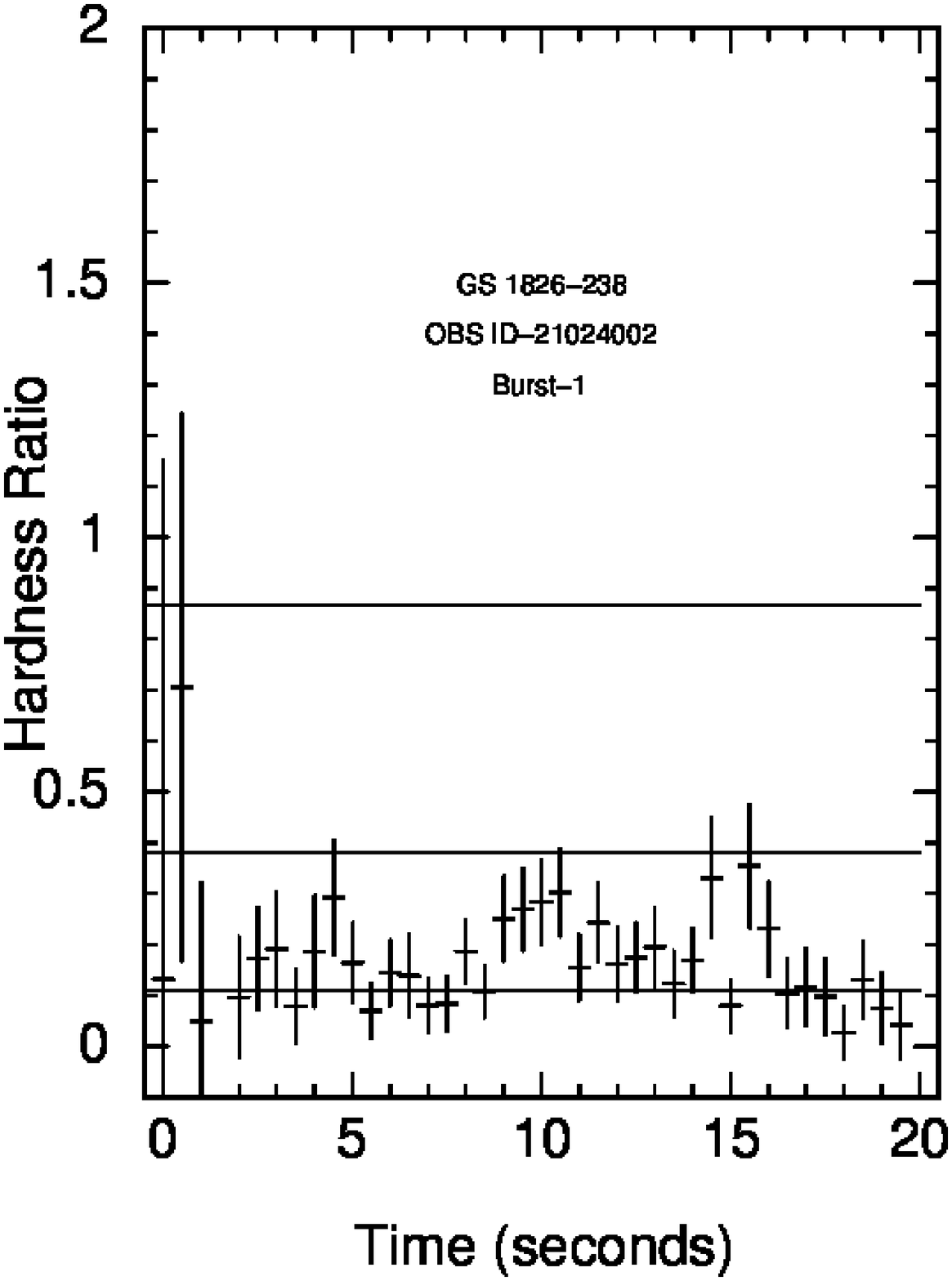}
\end{minipage}
\label{HR}
\end{figure*}

\begin{figure*}
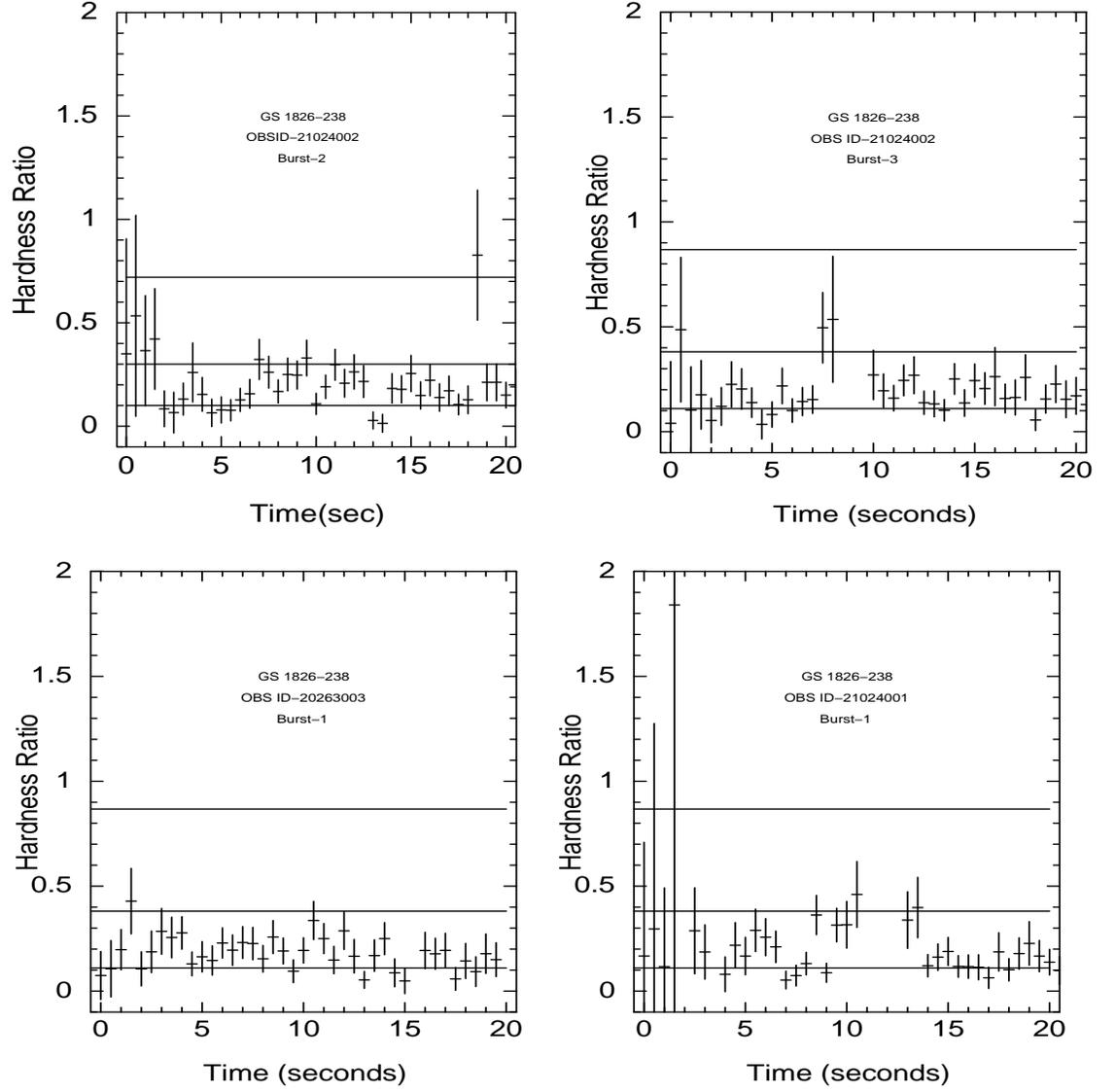

\begin{minipage}{0.43\textwidth}
\includegraphics[height=3.in,width=3.0in,angle=0]{fig-2s.ps}
\end{minipage}
\hspace{0.2\textwidth}
\begin{minipage}{0.43\textwidth}
\includegraphics[height=3.in,width=3.0in,angle=0]{fig-2t.ps}
\end{minipage}
\hspace{0.2\textwidth}
\begin{minipage}{0.35\textwidth}
\includegraphics[height=3.in,width=3.0in,angle=0]{fig-2u.ps}
\end{minipage}
\hspace{0.25\textwidth}
\begin{minipage}{0.4\textwidth}
\includegraphics[height=3.in,width=3.0in,angle=0]{fig-2v.ps}
\end{minipage}
 \caption{Hardness Ratio between two energy bands (1.8-10)~keV and (15.0-30.)~keV near the peak of bursts, obtained
 using background subtracted lightcurves from two instruments \emph{MECS} and \emph{PDS}. For details about background subtraction
 see text. The three horizontal lines starting from bottom in each plot represents ratios at 2.0~keV, 2.5~keV and 3.0~keV respectively.}
\label{HR}
\end{figure*}

\begin{figure*}
\begin{minipage}{0.35\textwidth}
\includegraphics[height=3.in,width=3.0in,angle=0]{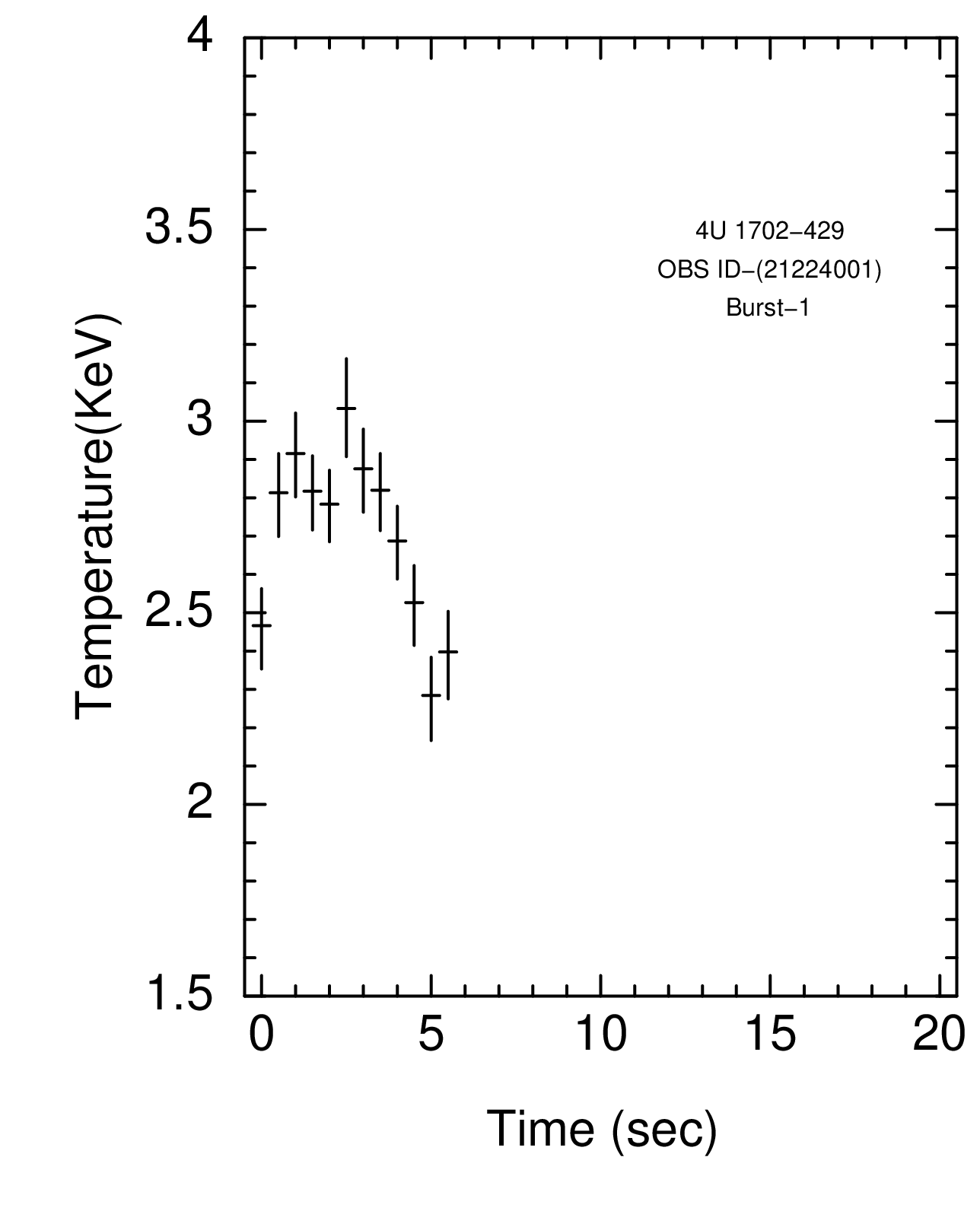}
\end{minipage}
\hspace{0.23\linewidth}
\begin{minipage}{0.35\textwidth}
\includegraphics[height=3.in,width=3.0in,angle=0]{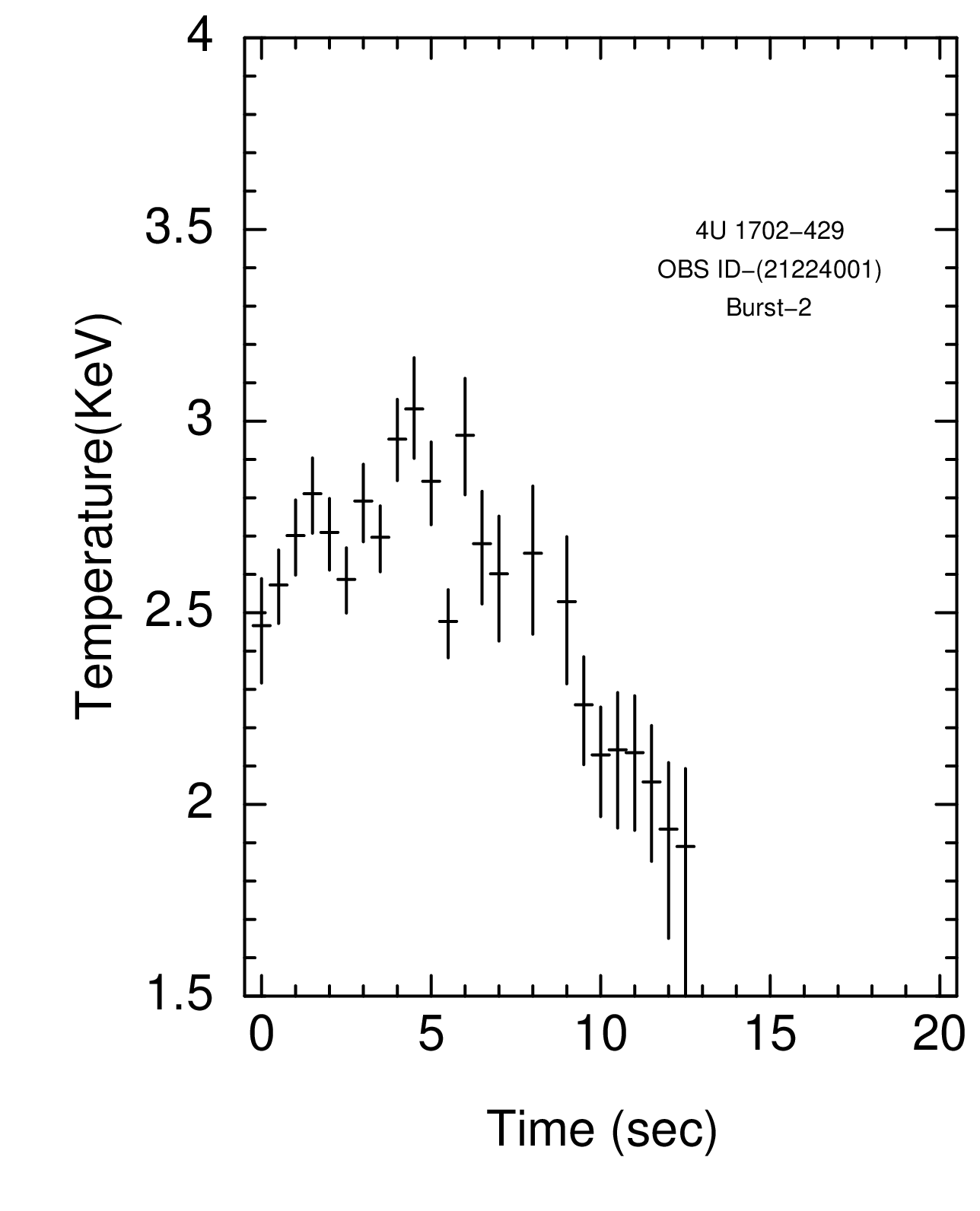}
\end{minipage}
\hspace{0.23\linewidth}
\begin{minipage}{0.4\textwidth}
\includegraphics[height=3.in,width=3.0in,angle=0]{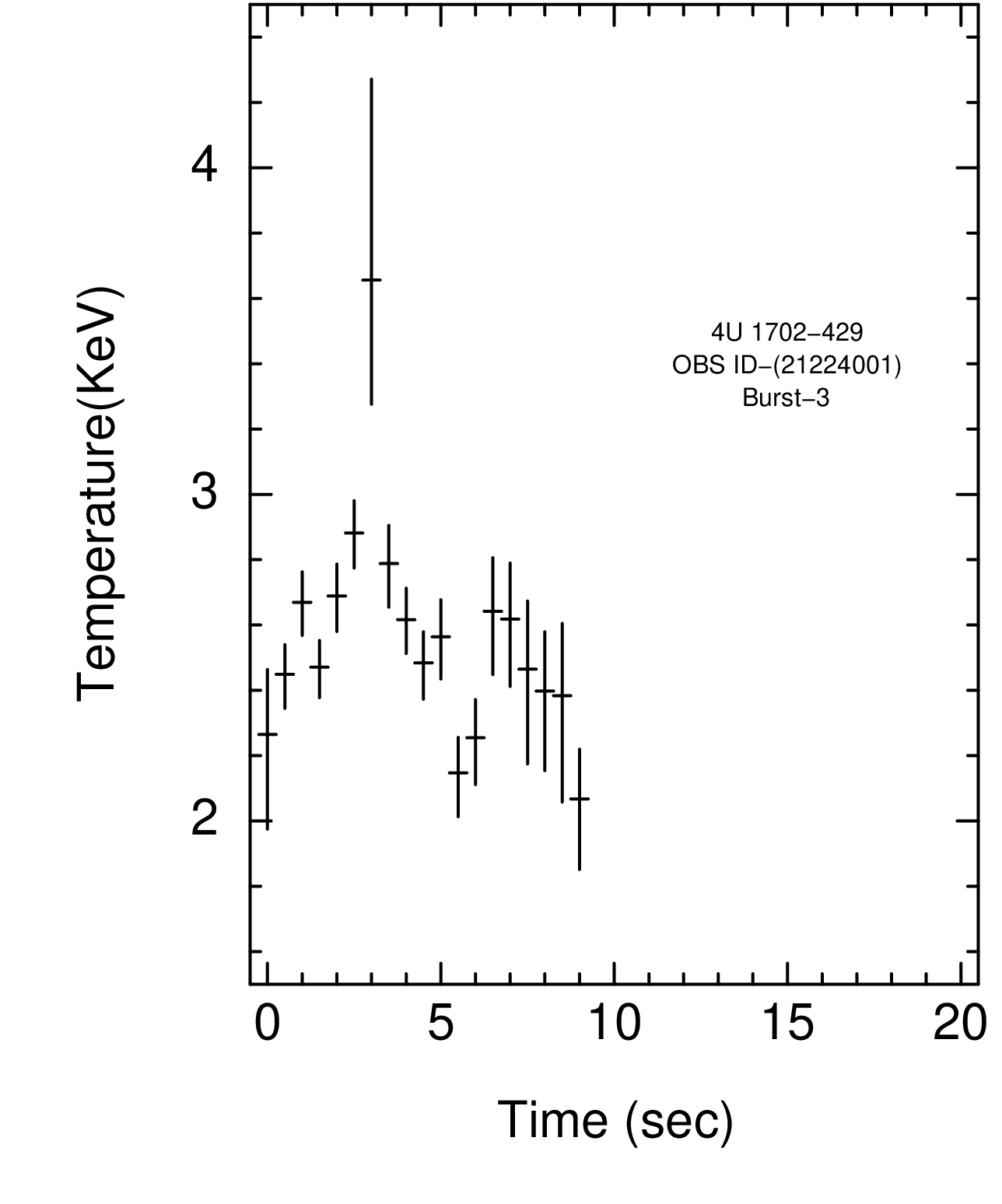}
\end{minipage}
\hspace{0.23\linewidth}
\begin{minipage}{0.4\textwidth}
\includegraphics[height=3.in,width=3.0in,angle=0]{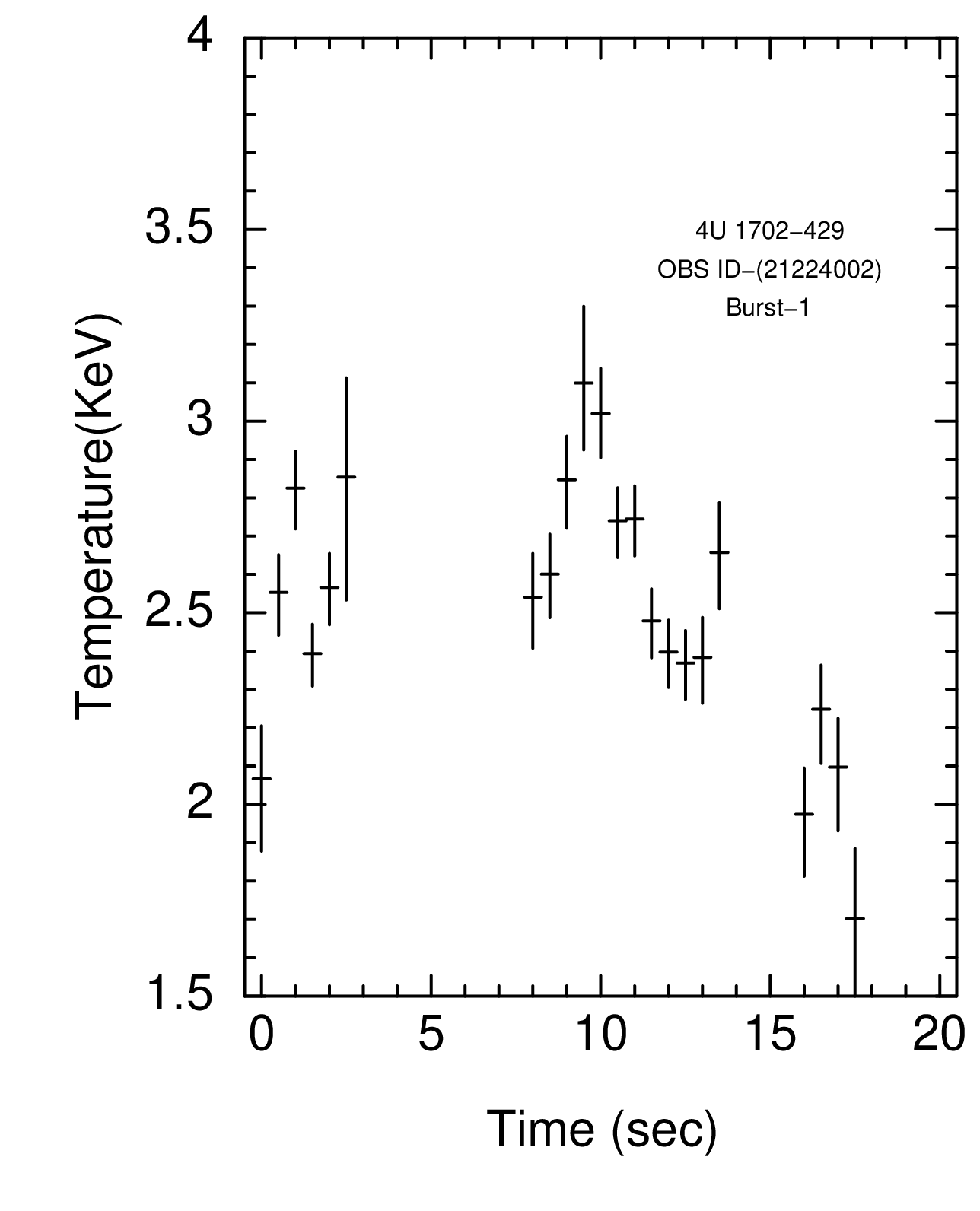}
\end{minipage}
\hspace{0.23\linewidth}
\begin{minipage}{0.35\textwidth}
\includegraphics[height=3.in,width=3.0in,angle=0]{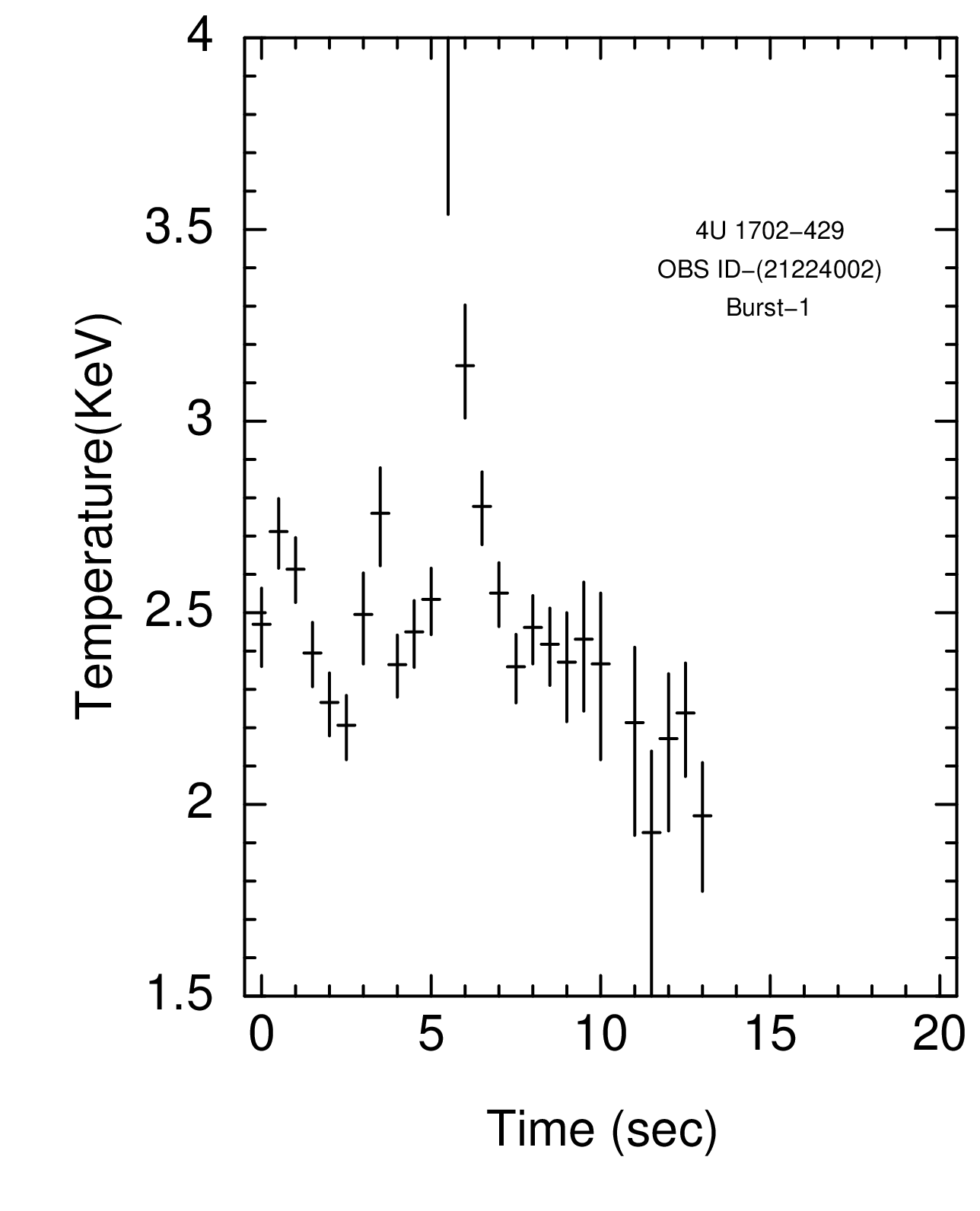}
\end{minipage}
\hspace{0.28\linewidth}
\begin{minipage}{0.35\textwidth}
\includegraphics[height=3.in,width=3.0in,angle=0]{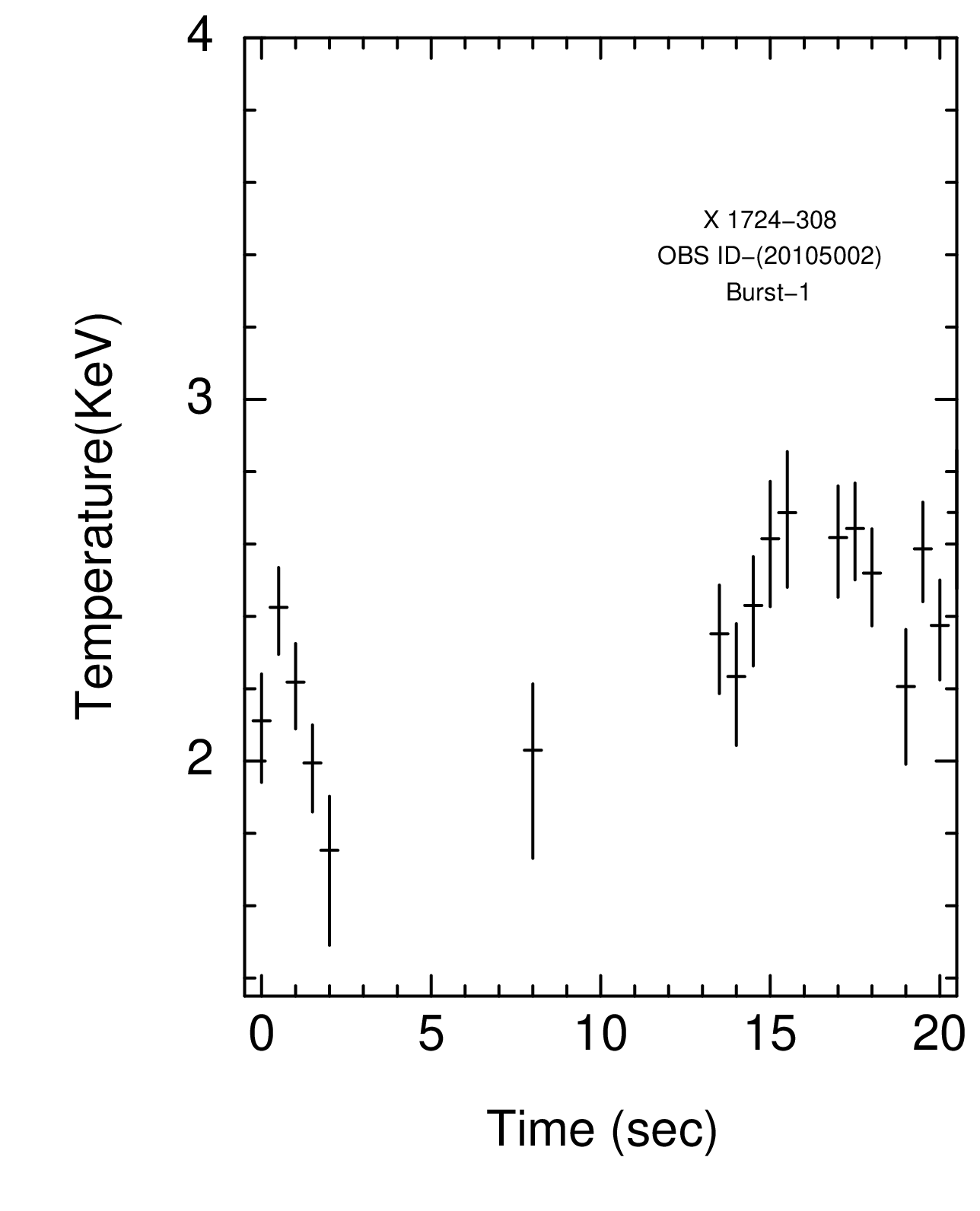}
\end{minipage}
\label{Temperature}
\end{figure*}

\begin{figure*}
\begin{minipage}{0.4\textwidth}
\includegraphics[height=3.in,width=3.0in,angle=0]{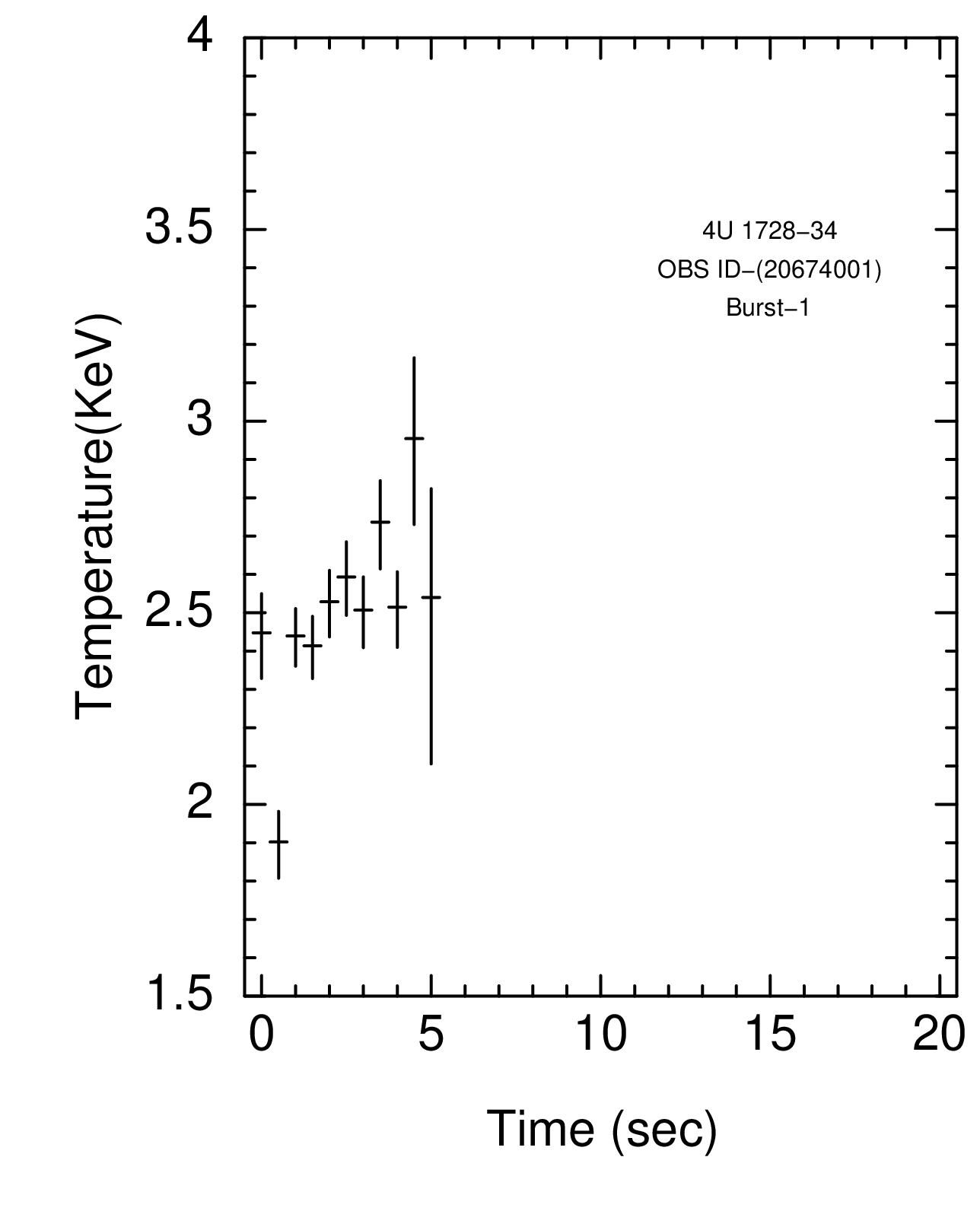}
\end{minipage}
\hspace{0.23\textwidth}
\begin{minipage}{0.4\textwidth}
\includegraphics[height=3.in,width=3.0in,angle=0]{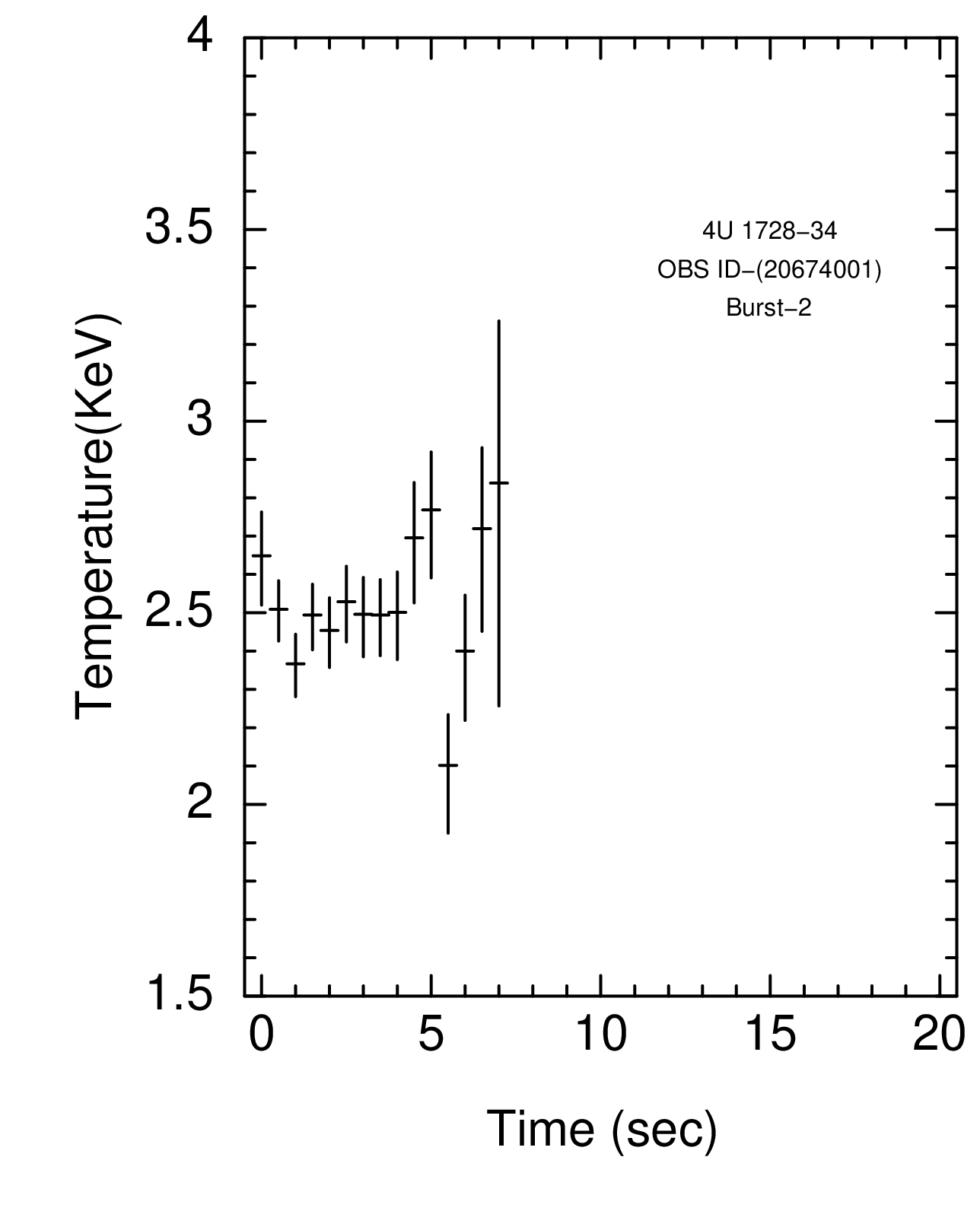}
\end{minipage}
\hspace{0.23\textwidth}
\begin{minipage}{0.35\textwidth}
\includegraphics[height=3.in,width=3.0in,angle=0]{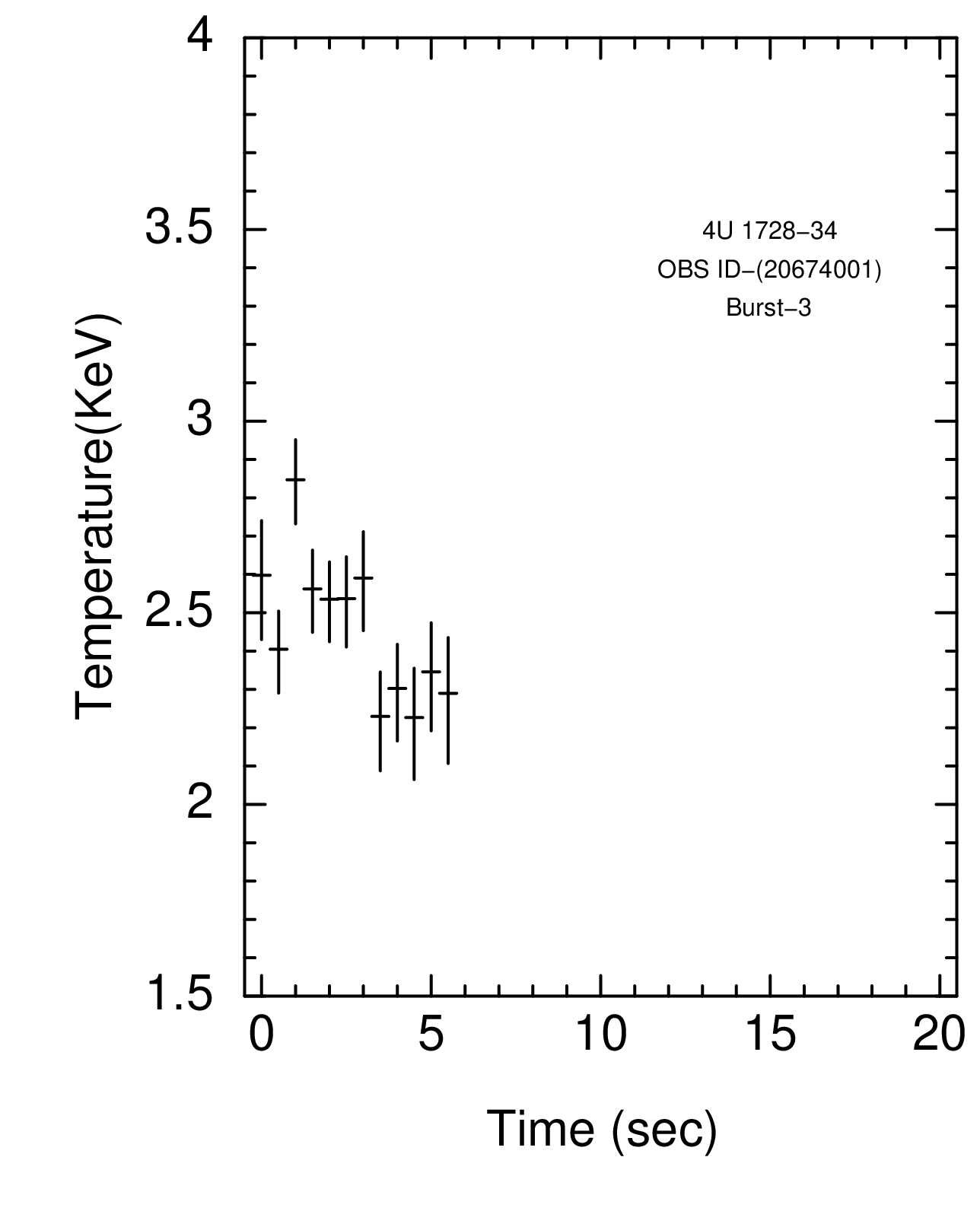}
\end{minipage}
\hspace{0.23\textwidth}
\begin{minipage}{0.35\textwidth}
\includegraphics[height=3.in,width=3.0in,angle=0]{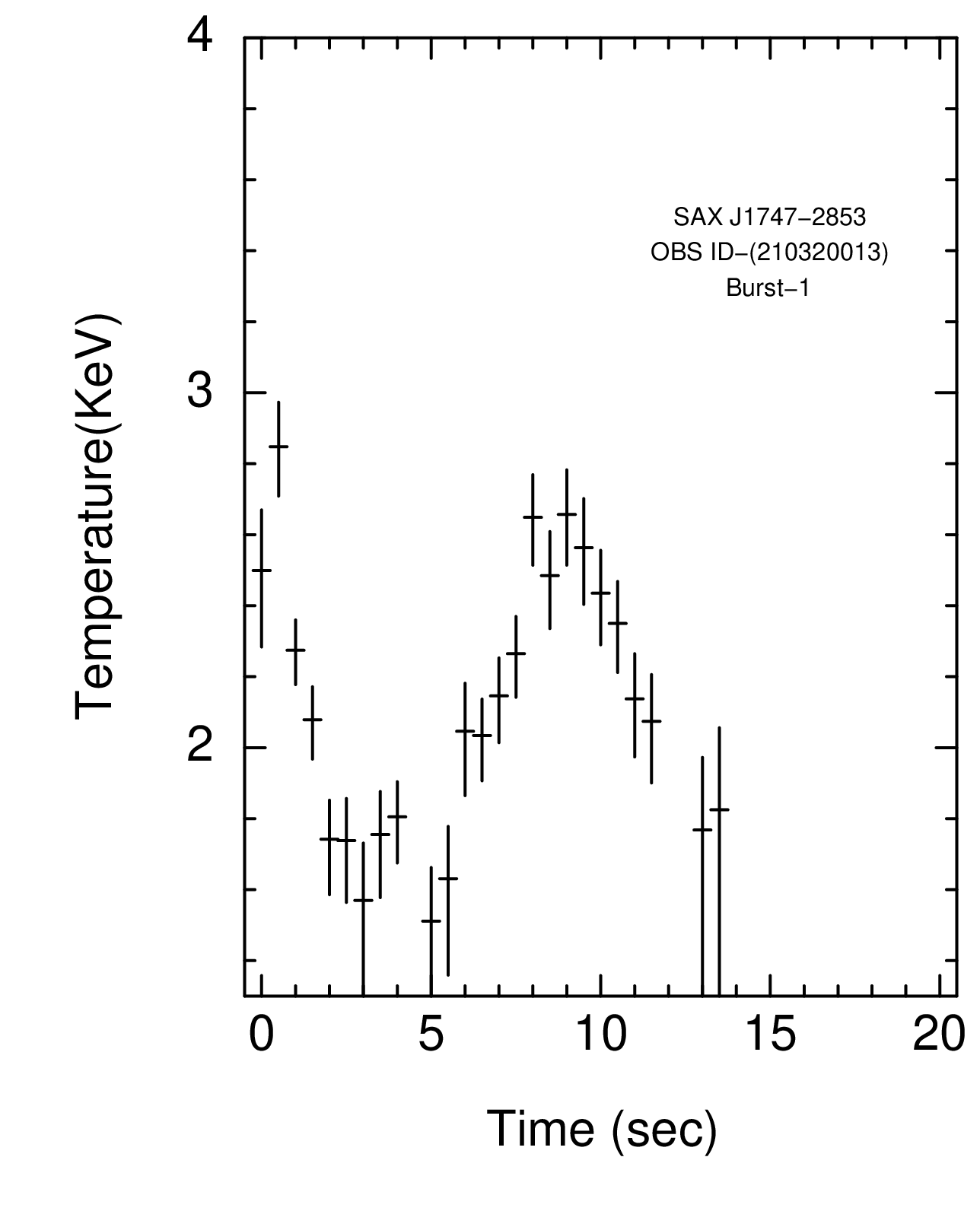}
\end{minipage}
\hspace{0.23\textwidth}
\begin{minipage}{0.4\textwidth}
\includegraphics[height=3.in,width=3.0in,angle=0]{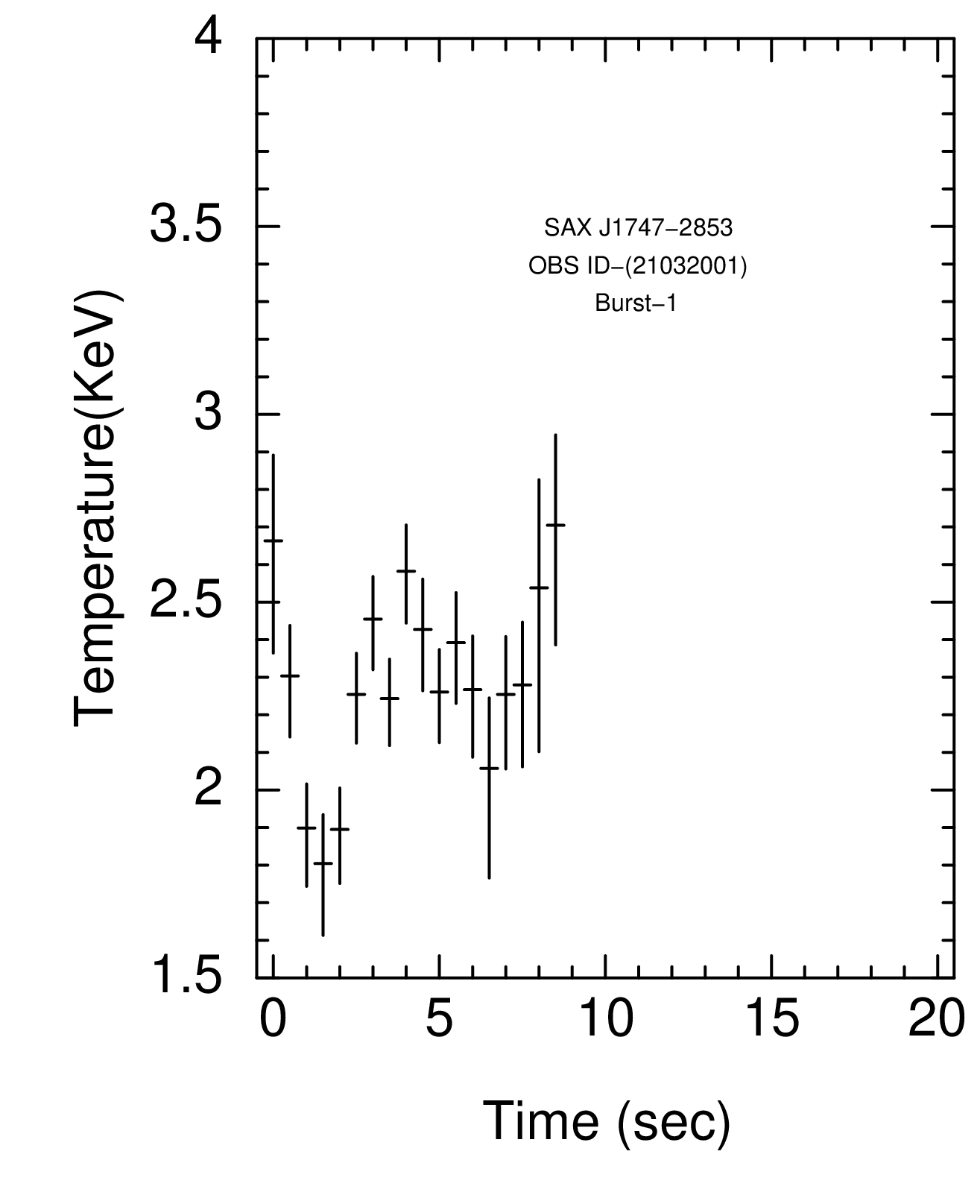}
\end{minipage}
\hspace{0.23\textwidth}
\begin{minipage}{0.4\textwidth}
\includegraphics[height=3.in,width=3.0in,angle=0]{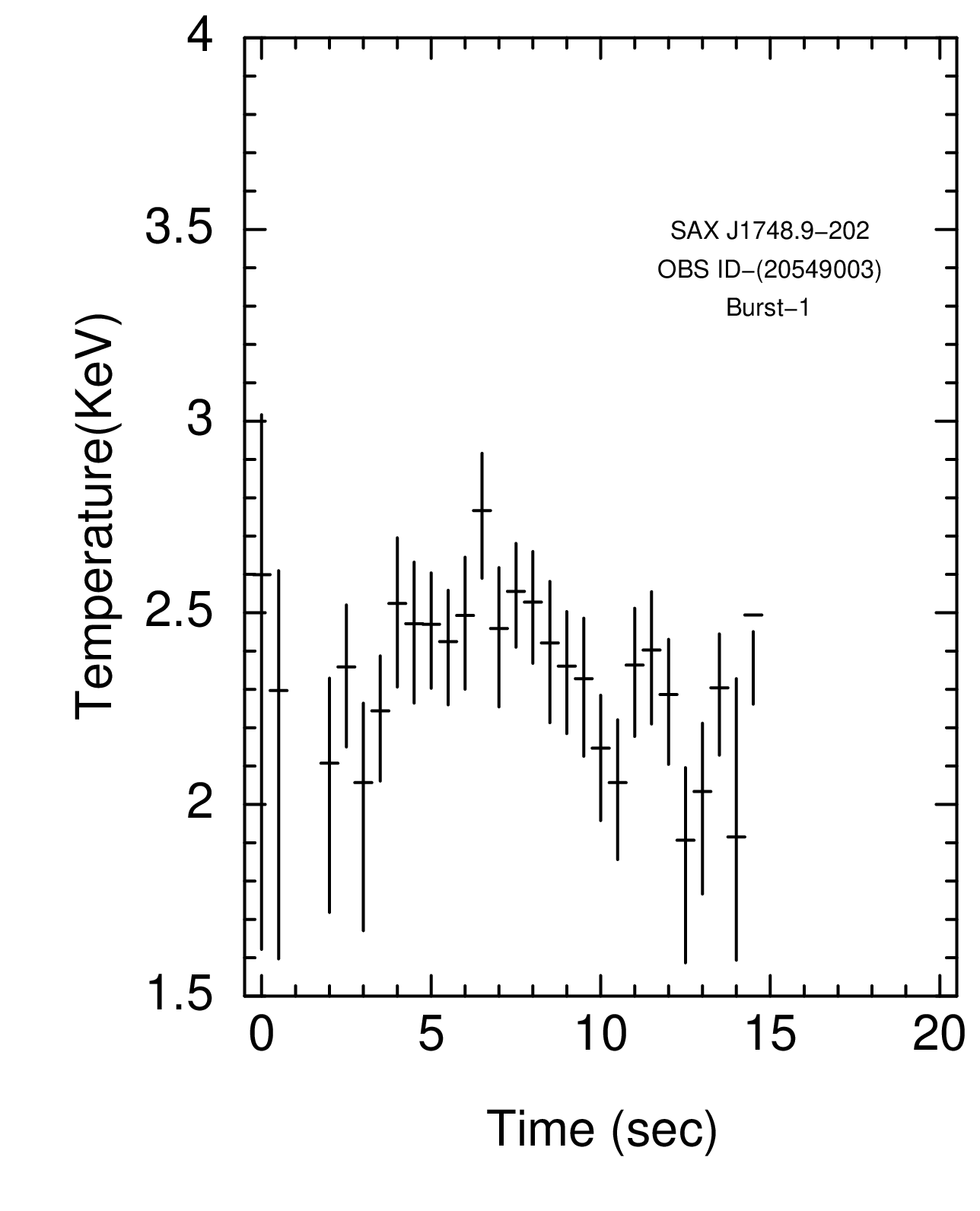}
\end{minipage}
\label{Temperature}
\end{figure*}

\begin{figure*}
\begin{minipage}{0.35\textwidth}
\includegraphics[height=3.in,width=3.0in,angle=0]{fig-4m.ps}
\end{minipage}
\hspace{0.23\textwidth}
\begin{minipage}{0.35\textwidth}
\includegraphics[height=3.in,width=3.0in,angle=0]{fig-4n.ps}
\end{minipage}
\hspace{0.23\textwidth}
\begin{minipage}{0.4\textwidth}
\includegraphics[height=3.in,width=3.0in,angle=0]{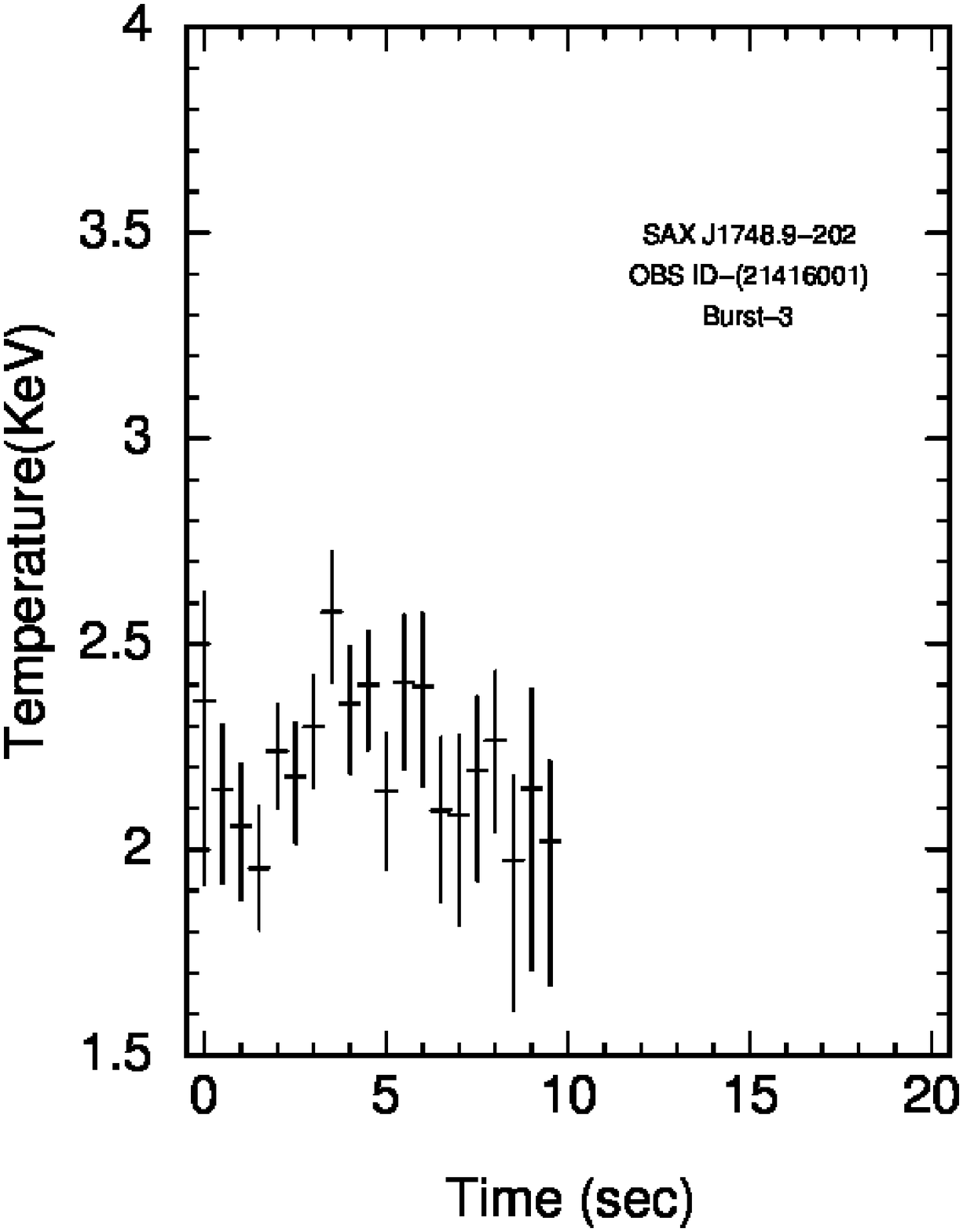}
\end{minipage}
\hspace{0.23\textwidth}
\begin{minipage}{0.4\textwidth}
\includegraphics[height=3.in,width=3.0in,angle=0]{fig-4p.ps}
\end{minipage}
\hspace{0.23\textwidth}
\begin{minipage}{0.35\textwidth}
\includegraphics[height=3.in,width=3.0in,angle=0]{fig-4q.ps}
\end{minipage}
\hspace{0.28\textwidth}
\begin{minipage}{0.35\textwidth}
\includegraphics[height=3.in,width=3.0in,angle=0]{fig-4r.ps}
\end{minipage}
\label{Temperature}
\end{figure*}

\begin{figure*}
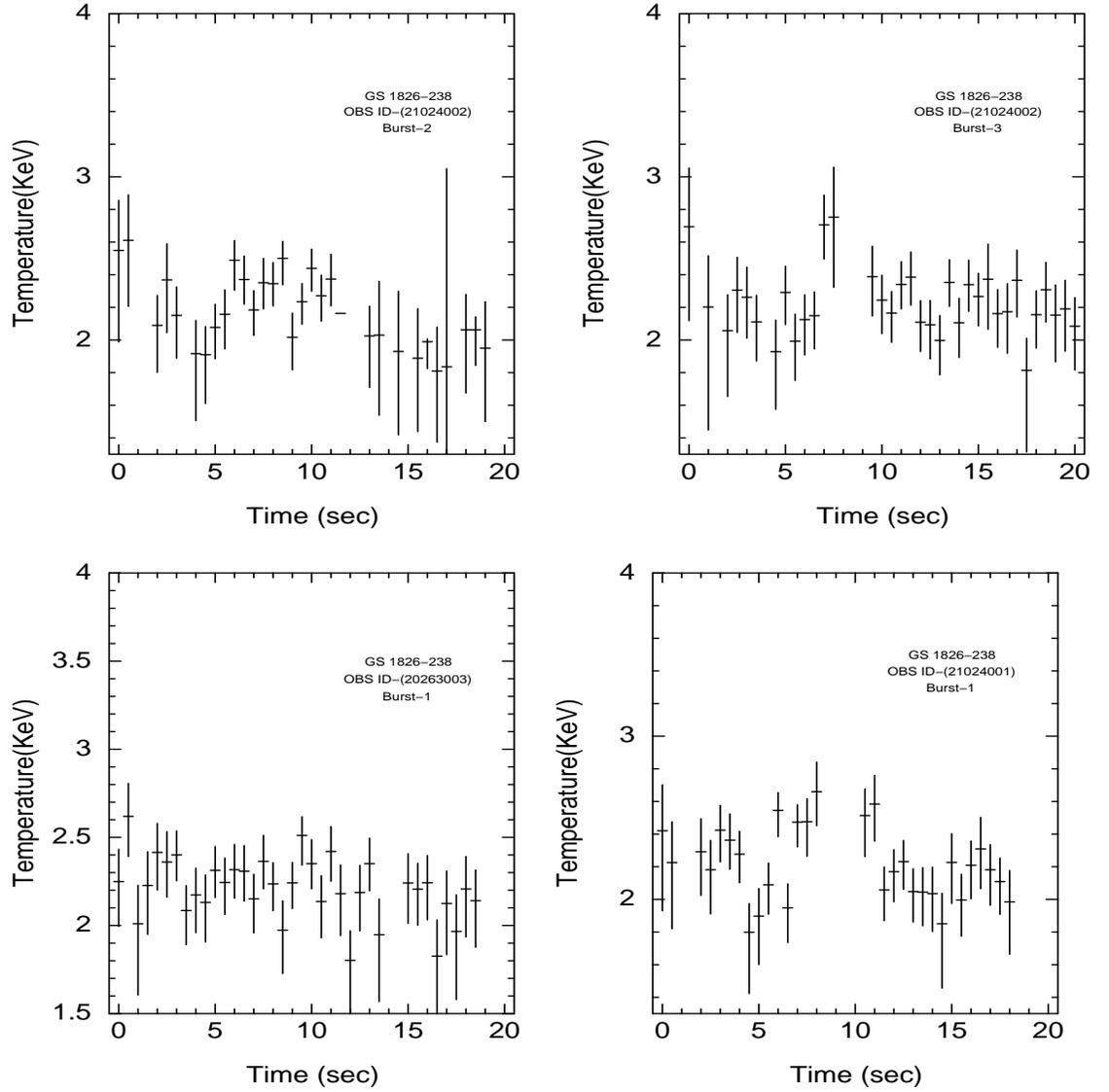

\begin{minipage}{0.4\textwidth}
\includegraphics[height=3.in,width=3.0in,angle=0]{fig-4s.ps}
\end{minipage}
\hspace{0.23\textwidth}
\begin{minipage}{0.4\textwidth}
\includegraphics[height=3.in,width=3.0in,angle=0]{fig-4t.ps}
\end{minipage}
\hspace{0.23\textwidth}
\begin{minipage}{0.35\textwidth}
 \includegraphics[height=3.in,width=3.0in,angle=0]{fig-4u.ps}
 \end{minipage}
 \hspace{0.25\textwidth}
 \begin{minipage}{0.4\textwidth}
 \includegraphics[height=3.in,width=3.0in,angle=0]{fig-4v.ps}
 \end{minipage}
 \caption{Temperature evolution near peak of bursts in different sources. 4U~1702--429 and 4U~1728--34,
 showing the values greater than 2.7~keV.}
 \label{Temperature}
\end{figure*}

\end{document}